\documentclass[11pt,a4paper]{article}

\usepackage[margin=1in]{geometry}  
\usepackage{graphicx}              
\usepackage{amsmath,amsfonts,amsthm,amssymb,mathrsfs}               
\usepackage{enumerate}
\usepackage{url}
\usepackage{breakurl}
\usepackage{tikz}
\usepackage{IEEEtrantools}
\usepackage{subfigure}
\usepackage[numbers,sort&compress]{natbib}
\usepackage[para]{footmisc}

\usepackage{paralist}
\usepackage[toc]{appendix}

\usetikzlibrary{automata}
\usetikzlibrary{positioning}
\DeclareMathOperator*{\argmin}{arg\,min}
\allowdisplaybreaks

\newtheorem{thm}{Theorem}[]

\newtheorem{rmk}[thm]{Remark}

\title{Characterizing Information Spreading in Online Social Networks}

\author{Sai Zhang\footnotemark[2]
\and
Ke Xu\footnotemark[3]
\and
Xi Chen\footnotemark[4]
\and
Xue Liu\footnotemark[4]}

\renewcommand{\thefootnote}{\fnsymbol{footnote}}

\footnotetext[2]{Institute for Interdisciplinary Information Sciences, Tsinghua University, Beijing, 100084, China. Email: \url{zhangsai13@mails.tsinghua.edu.cn}.}
\footnotetext[3]{Department of Computer Science and Technology, Tsinghua University, Beijing, 100084, China. Email: \url{xuke@mail.tsinghua.edu.cn}.}
\footnotetext[4]{School of Computer Science, McGill University, Montreal, QC H3A 0E9, Canada. Email: \url{{xchen100,xueliu}@cs.mcgill.ca}.}

\date{}

\begin{document}


\maketitle

\begin{abstract}
Online social networks (OSNs) are changing the way in which the information spreads throughout the Internet. A deep understanding of the information spreading in OSNs leads to both social and commercial benefits. In this paper, we characterize the dynamic of information spreading ({e.g.}, how fast and widely the information spreads against time) in OSNs by developing a general and accurate model based on the Interactive Markov Chains (IMCs) and mean-field theory. This model explicitly reveals the impacts of the network topology on information spreading in OSNs. Further, we extend our model to feature the time-varying user behaviors and the ever-changing information popularity. The complicated dynamic patterns of information spreading are captured by our model using \emph{six} key parameters. Extensive tests based on Renren's dataset validate the accuracy of our model, which demonstrate that it can characterize the dynamic patterns of video sharing in Renren precisely and predict future spreading tendency successfully.
\end{abstract}
\clearpage
\section{Introduction}
\renewcommand{\thefootnote}{\arabic{footnote}}
With fast developments of the Internet and Web 2.0, online social networks (OSNs) generated by various Web applications are playing important roles in the information spreading throughout the Internet. Getting, publishing and sharing information with forums, video-sharing sites (VSSes) and social networking sites (SNSes) are becoming increasingly popular. A recent report~\cite{fn1} shows that as of May 2013, almost 72\% online U.S. adults use SNSes, up from 67\% in late 2012 and only 8\% in February 2005. In August 2013~\cite{fn2}, there were about 500 million tweets per day on Twitter and a peak of 143,199 tweets per second was observed during the airing of \emph{Castle in the Sky}~\cite{fn3}. SNSes become so popular that they begin to influence users' purchase decisions. Statistics~\cite{fn4} show that 74\% consumers make their purchase decisions based on SNSes. In addition, in emergency situations such as fires and earthquakes, information spreading in SNSes can provide valuable knowledge that is crucial in life-saving. Approximately 96\% earthquakes of Japan Meteorological Agency (JMA) seismic intensity scale 3 or more are detected merely by monitoring tweets \cite{Sakaki2010}. Moreover, SNSes even facilitate the mobilization of mass movements~\cite{fn5}. Therefore, it is important and necessary to study the nature of information spreading in OSNs to promote viral marketing, improve social benefits, and maintain social security in the practical aspects, as well as understand the characteristics of complex networks and individual behaviors in the scientific aspects.

Numerous works \cite{Galuba2010,Guille2012,Yang2010,Leskovec2007,Moreno2004,Barthelemy2005,Weng2014,Cheng2014} have studied various spreading processes in networks. Attributed to the complexity of the underlying network topology and the diversity of collective behaviors, these models only focus on simplified scenarios where several important factors (e.g., the network structure and the spreading variation) of the spreading process are missing. In particular, several works \cite{Leskovec2007,Yang2010} are limited by the fact that they have ignored the network topology and only modelled the spreading process in a global perspective, while other approaches \cite{Galuba2010,Guille2012,Cheng2014} which consider the network structure cannot work when the network is implicit or even unknown. On the other hand, various studies \cite{Kleinberg2002,Vlachos2004,Benevenuto2009,Yang2011} have shown that the temporal dynamic of online media exhibits great burstiness, diurnal patterns and periodicity, which cannot be fully captured by those ``static'' models~\cite{Galuba2010,Leskovec2007,Moreno2004,Barthelemy2005}.

To tackle the aforementioned issues, in this paper, we first develop a naive model starting from a general mechanism of the information spreading in realistic OSNs, which facilitates us to study the impacts of the underlying network topology on the spreading process. We mainly consider two key measurements of the network topology, namely, the degree distributions (the first-order correlation) and the degree-degree correlations (the second-order correlation), which balances the global-local paradox. Our extensive theoretical analysis accurately reveals the relations between the network heterogeneity and the information spreading, and further demonstrates that the positive degree-degree correlations \emph{do not} always inhibit the spreading process. This relatively negative result has ever been misunderstood by previous studies~\cite{Moreno2003,Nekovee2007}.
Furthermore, we extend our naive model and equip it with time-varying parameters, which characterize the user behavior and information popularity in time domain. Our analysis on this extended model shows that the temporal variation patterns of the information spreading in OSNs are approximately determined by \emph{six} parameters. To our knowledge, this is \emph{the first spreading model that characterizes the dynamic patterns of the information spreading in realistic complex networks in closed forms}. By testing on a data set of video sharing in Renren~\cite{fn6}, one of the largest SNSes in China, we validate that our extended model can depict the spreading dynamic precisely. In particular, we verify that the time-varying model can successfully predict the long-range spreading tendency given history data.

The rest of the paper is structured as follows. Section 2 sums up the terminologies and notations used in this paper. Section 3 describes our naive probabilistic model and then gives the deterministic dynamical system derived from the model. In Section 4, based on the naive model, we comprehensively study the impacts of the underlying network topology on the information spreading in both uncorrelated and correlated heterogeneous networks. The extended spreading model with time-varying parameters and its analysis are given in Section 5. Section 6 carries out several numerical studies to verify previous theoretical conclusions, and uses the Levenberg-Marquardt (LM) algorithm \cite{Levenberg1944} to learn the dynamic patterns of video sharing in Renren based on the time-varying model, which can be further used to predict the future tendency of video sharing given history data. Finally, we summarize related work in Section 7, conclude this paper and discuss our future work in Section 8.

\section{Preliminaries}
In this section, we list several common terminologies and notations in network science, which are frequently used in this paper.

The \emph{degree} \cite{Newman2010} of a vertex in an undirected network is defined by the the number of edges attached to it. We denote the largest degree in a network as $k_c$ and always use lower-case $k$ to denote degrees. The \emph{degree distribution} $P(k)$ \cite{Newman2010} describes the fraction of vertices with degree $k$. We denote the average degree as $\langle k\rangle$, which equals to $\sum_k kP(k)$. Similarly, the average square degree $\langle k^2\rangle$ equals to $\sum_k k^2P(k)$. Note that we always approximate the summation of sequence by its integral to simplify computations, e.g., $\sum kP(k)\approx\int kP(k)$, which is a standard approximation method in network science.

Degree distribution characterizes the network topology from a local perspective, and different networks with the same degree distribution share similar topologies. Here, we consider the scale-free (SF) networks, the most common networks in real world \cite{Newman2010}. The degree distribution of an SF network follows the power-law distribution. Formally speaking, in an SF network, the degree distribution $P(k)$ is defined by \cite{Newman2010}
\begin{equation}
P(k)=
\begin{cases}
\displaystyle Zk^{-\gamma} & \text{if }k\geq m,\\
\displaystyle 0&\text{otherwise},
\end{cases}
\label{eq:1}
\end{equation}
where $m$ is the minimum degree of the network and is set to be 1 in this paper, and the normalization constant $Z$ approximately equals to $(\gamma-1)m^{\gamma-1}$. In order to ensure finite average degree in the SF network, we shall keep $\gamma>2$. Particularly, the case of $\gamma=3$ is the degree distribution of the networks generated by the Barab\'asi-Albert (BA) model \cite{Barabasi1999}.

A network is \emph{heterogeneous} \cite{Barthelemy2005} if there are vertices whose degrees are significantly larger than the average degree. Otherwise, the network is called \emph{homogeneous} \cite{Barthelemy2005}. More precisely, we can use $\langle k^2\rangle/\langle k\rangle$ to quantify the network heterogeneity approximately, {i.e.}, a network usually becomes more heterogeneous if its $\langle k^2\rangle/\langle k\rangle$ gets larger. Thus, heterogeneity characterizes the irregularity of the network topology from the degree distribution perspective. It is not hard to find that in an SF network, as the exponent $\gamma$ gets smaller, the network becomes more heterogeneous (large degree appears with higher probability).

Most realistic networks exhibit non-trivial connectivity patterns \cite{Newman2010}. For instance, most OSNs present \emph{assortativity} \cite{Mislove2007,Chun2008,Jiang2010}, while the Web networks and the Internet are always \emph{disassortative} \cite{Newman2002}. These connectivity trends among vertices with different degrees cannot be described by degree distributions. Actually, various connectivity patterns influence the network topology drastically and result in different degree distributions.

Degree distribution is often referred as the \emph{first-order correlation} \cite{Bogua2003}. To model the above connectivities, we can define the conditional probability $P(k',k'',...,k^{(n)}\mid k)$, the probability of vertex with degree $k$ attaching simultaneously to other $n$ vertices with corresponding degrees $k',k'',...,k^{(n)}$, which characterizes higher-order (the \emph{$n$th-order}) correlation. The network is referred as \emph{uncorrelated} \cite{Bogua2003} if there is no conditional probability in the network connectivity. Otherwise, the network is \emph{correlated} \cite{Bogua2003}. It is known that in Markovian networks \cite{Boguna2002}, which is the main concern in the following sections, only the degree distribution $P(k)$ and the degree-degree correlations $P(k'\mid k)$ are taken into account. If we define the joint probability distribution $P(k,k')$ by the probability of randomly selecting an edge which connects a vertex with degree $k$ and another vertex with degree $k'$ simultaneously \cite{Newman2010}, and define the \emph{excess degree distribution} $q(k)\triangleq\sum_{k'}P(k,k')$ which equals to $kP(k)/\langle k\rangle$ \cite{Newman2010}, we have
\begin{equation}
P(k,k')=q(k)P(k'\mid k).\label{3}
\end{equation}
Therefore, the joint distribution $P(k,k')$ totally determines the topology of the Markovian network. In addition, we have $P(k,k')=q(k)q(k')$ in uncorrelated networks, which leads to $P(k'\mid k)=q(k')$. These results are frequently used in the following sections. Another measurement which describes the second-order correlation is the \emph{average nearest neighbors degree} (ANND) $\langle k_{nn}\rangle\triangleq\sum_{k'}k'P(k'\mid k)$ \cite{Pastor-Satorras2001}, {i.e.}, the average neighbors degree of the vertex with degree $k$. In uncorrelated networks, we have $\langle k_{nn}\rangle=\langle k^2\rangle/\langle k\rangle$ which is independent on $k$.

\section{Naive Model}
We consider an undirected network $N=(V,E)$, where $V$ denotes the set of vertices and $|V|=n$, and $E$ denotes the set of edges. Note that we sometimes refer to vertices as users in OSNs. Information $\mathcal{I}$ ({e.g.}, a message, photo or video) spreads along its edges. We first divide the states of vertices into three kinds: \emph{ignorant}, \emph{active} and \emph{indifferent}. Ignorant vertices are those who are unaware of $\mathcal{I}$, active vertices are those who have noticed $\mathcal{I}$ and then propagate ({e.g.}, forwarding or sharing in SNSes) it, and indifferent vertices are those who are aware of it but do not propagate it. Since active vertices have propagated $\mathcal{I}$, their ignorant neighbors can be activated with a certain probability. Based on the fact that active vertices can only affect their neighbors in a finite time duration due to the popularity decay, we assume that active vertices become \emph{quiet} (a new state in which they get no influence on their neighbors any more) with a certain probability spontaneously.

\begin{figure}[!tpb]
\centering
\subfigure[Full states]
{
\scalebox{0.8}{
\begin{tikzpicture}[minimum size=0.1mm,node distance=1.5cm,semithick,auto]
\node[state] (i)              {$i$};
\node[state] (r) [right=of i] {$r$};
\node[state] (a) [below=of i] {$a$};
\node[state] (q) [right=of a] {$q$};
\node[state] (f) at (4,-1.25) {$\bullet$};
\path[->]
    (i) edge [loop above] node {$1-\lambda$} (i)
    (a) edge [loop below] node {$1-\beta$} (a)
    (i) edge              node {$(1-\alpha)\lambda$} (r)
    (i) edge              node {$\alpha\lambda$} (a)
    (a) edge              node {$\beta$} (q)
    (r) edge              node [above right] {$1$} (f)
    (q) edge              node [below right] {$1$} (f);
\end{tikzpicture}
\label{fig:11}
}
}
\hspace{.3in}
\subfigure[Simplified states]
{
\scalebox{0.8}{
\begin{tikzpicture}[minimum size=0.1mm,node distance=1.5cm,semithick,auto]
\node[state] (i)              {$i$};
\node[state] (a) [below=of i] {$a$};
\node[state] (f) at (2.2,-1.25) {$\bullet$};
\path[->]
    (i) edge [loop above] node {$1-\lambda$} (i)
    (a) edge [loop below] node {$1-\beta$} (a)
    (i) edge              node [above right] {$(1-\alpha)\lambda$} (f)
    (i) edge              node {$\alpha\lambda$} (a)
    (a) edge              node [below right] {$\beta$} (f);;
\end{tikzpicture}
\label{fig:12}
}
}
\caption{State transition diagram of the naive model. (a) The diagram with fully-defined vertex states. (b) The diagram with simplified vertex states. $i$, $a$, $r$, $q$ are short for \emph{ignorant}, \emph{active}, \emph{indifferent} and \emph{quiet}, respectively.}
\label{fig:1}
\end{figure}
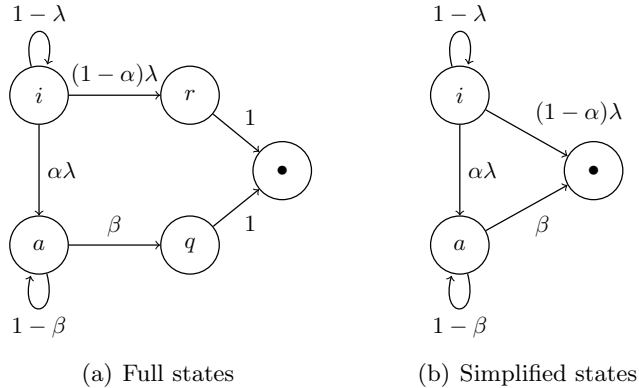

Figure~\ref{fig:11} illustrates our naive model. In network $N$, there are precisely four kinds of vertices, i.e., \emph{ignorant}, \emph{active}, \emph{indifferent} and \emph{quiet} vertices. We assume that information $\mathcal{I}$ spreads in $N$ based on the following mechanisms:
\begin{enumerate}[(i)]
\itemsep=-2pt
\item While keeping their states unchanged with probability $1-\lambda$, ignorant vertices \emph{with active neighbors} will notice $\mathcal{I}$ to become active or indifferent with probability $\lambda$.
\item Vertices who have noticed $\mathcal{I}$  and are as yet inactive will become active with probability $\alpha$ or indifferent with probability $1-\alpha$.
\item Active vertices become quiet spontaneously with probability $\beta$.
\end{enumerate}

From the above rules, we find that indifference and quietness are two final states of vertices. In fact, the process of information spreading is also the process of vertex state decay with indifference and quietness as its final states. To simplify the state transition diagram, here we add another state ``$\bullet$'' to uniformly describe all vertices aware of $\mathcal{I}$. Since indifferent and quiet vertices both belong to this kind of vertices and cannot affect their neighbors, they become ``$\bullet$'' absolutely (see Fig.~\ref{fig:11}). The simplified state transition diagram is shown in Fig.~\ref{fig:12}. Note that the spreading process terminates when there is no active vertex in the network.

We have shown that each vertex in $N$ changes its state based on the state transition diagram and its neighbors' states. This process is what the \emph{Interactive Markov Chains} (IMCs) \cite{Conlisk1976} deal with. Therefore, by keeping the above spreading mechanisms in mind, we can derive (details are given in Appendix A) the following set of coupled differential equations (also a non-linear dynamical system) that characterize the dynamic of the spreading process on a mean-field level:
\begin{IEEEeqnarray}{rCl}
\frac{\mathrm{d} i_k(t)}{\mathrm{d}t}&=&-\lambda k i_k(t)\sum_{k'}P(k'\mid k)a_{k'}(t),\IEEEyesnumber\IEEEyessubnumber\label{eq:0a}\\
\frac{\mathrm{d} a_k(t)}{\mathrm{d}t}&=&\alpha\lambda k i_k(t)\sum_{k'}P(k'\mid k)a_{k'}(t)-\beta a_k(t),\IEEEyessubnumber\label{eq:0b}\\
\frac{\mathrm{d} r_k(t)}{\mathrm{d}t}&=&(1-\alpha)\lambda k i_k(t)\sum_{k'}P(k'\mid k)a_{k'}(t),\IEEEyessubnumber\label{eq:0c}\\
\frac{\mathrm{d} q_k(t)}{\mathrm{d}t}&=&\beta a_k(t),\IEEEyessubnumber\label{eq:0d}
\end{IEEEeqnarray}
where $i_k(t)$, $a_k(t)$, $r_k(t)$, $q_k(t)$ denote the fractions of vertices with degree $k$ in the states of ignorant, active, indifferent and quiet at time $t$, respectively, and integer $k'$ ranges from $0$ to $k_c$, which is the largest degree\footnote{We always make $k_c$ approach infinity in certain cases. We will see this approximation is reasonable as well as computationally helpful in the following sections.} in network $N$. Obviously, we have $i_k(t)+a_k(t)+r_k(t)+q_k(t)=1$. If we denote $i(t)$, $a(t)$, $r(t)$, $q(t)$ as the vertex fractions of the four states respectively, we also have $i(t)+a(t)+r(t)+q(t)=1$. In addition, we assume that at $t=0$, the initial conditions of the system are $i(0)=(n-1)/n\approx 1$, $a(0)=1/n\approx 0$, $r(0)=0$ and $q(0)=0$, where ``$\approx$'' makes sense if $n$ is large enough. Note that the defined network $N$ is a kind of \emph{undirected Markovian random network} \cite{Boguna2002}, whose topology is completely determined by the degree distribution $P(k)$ and the conditional probability $P(k'\mid k)$. In other words, we assume in approximation that all higher-order correlation functions can be obtained by random combinations of $P(k)$ and $P(k'\mid k)$.

\subsection{Three Indices Quantifying Information Spreading}
\newtheorem{definition}{\textbf{Definition}}
As $t\rightarrow \infty$, the process of information spreading reaches its equilibrium when there is no active vertices in the network. What matters most is the condition under which information $\mathcal{I}$ spreads out, as well as the range and the velocity of the spreading process. We now define several indices to quantify these attributes.
\begin{definition}
The spreading \textbf{prevalence} $\mathcal{P}$ is defined by $r_{\infty}+q_{\infty}$, where $r_{\infty}\triangleq\lim_{t\rightarrow\infty}r(t)$ and $q_{\infty}\triangleq\lim_{t\rightarrow\infty}q(t)$.
\end{definition}

The spreading prevalence $\mathcal{P}$ characterizes the final spreading range, {i.e.}, the fraction of vertices aware of the information when the spreading process ends. It is obvious that $0<\mathcal{P}\leq1$.

Consider the model parameter function $\rho(\alpha,\beta,\gamma)=\alpha\gamma/\beta$, which is denoted as $\rho$ for short. In fact, $\rho$ reflects the spreading potential of the information. Concretely, if the vertices in network $N$ get and propagate information $\mathcal{I}$ easily (correspond to large $\lambda$ and $\alpha$, respectively), and meanwhile, $\mathcal{I}$ loses its popularity slowly (corresponds to small $\beta$), then $\mathcal{I}$ will spread in a wide range. Otherwise, $\mathcal{I}$ can be trapped in a narrow area.
\begin{definition}
If there is a real number $\rho_c\geq0$ satisfying $\mathcal{P}>0$ if $\rho>\rho_c$, we refer to $\rho_c$ as the spreading \textbf{threshold}.
\end{definition}

If a function $f(t)$ evolves with a dominant term $e^{t/\tau}$, it is clear that the time scale $\tau$ controls the evolution velocity of $f$, {i.e.}, the smaller $\tau$ is, the faster $f$ evolves. In order to describe the velocity of the spreading process, we define the spreading efficiency as follows.
\begin{definition}
\label{def3}
The spreading \textbf{efficiency} $\mathcal{E}$ is defined by the reciprocal of the growth time scale $\tau$ in the information spreading, {i.e.}, $\mathcal{E}\triangleq 1/\tau$.
\end{definition}

Note that in Definition~\ref{def3} we treat the spreading process as the evolution process of $a(t)$. We will clarify this definition with the help of some analytic results derived in Section 4.

\section{Impacts of Network Topology on Information Spreading}
In this section, we study the influence of the underlying network topology on the spreading process in both uncorrelated and correlated heterogeneous networks, based on the naive model proposed in Section 3. Our philosophy is to study simple networks first, integrate the higher-order correlation gradually, and investigate how the added correlation influences the spreading process.
\subsection{Uncorrelated Heterogeneous Networks}
Here, we mainly concern how the degree distribution (the first-order correlation) affects the spreading dynamic. To clarify this, we focus on the SF networks and analyze the exponent $\gamma$'s function on spreading. We have the following theorems.
\newtheorem{theorem}{\textbf{Theorem}}
\begin{theorem}
\label{thm1}
In uncorrelated networks, the spreading threshold $\rho_c$ equals to ${\langle k\rangle}/{\langle k^2\rangle}$. In other words, the prevalence $\mathcal{P}>0$ if and only if $\rho>{\langle k\rangle}/{\langle k^2\rangle}$. In particular, in SF networks, we have
\begin{equation}
\label{eq:2}
\rho_c=
\begin{cases}
\displaystyle\frac{\gamma-3}{\gamma-2}&\text{if }\gamma>3,\\
\displaystyle 0&\text{if }2<\gamma\leq 3.\\
\end{cases}
\end{equation}
\end{theorem}
\begin{IEEEproof}
See Appendix B.
\end{IEEEproof}

\begin{rmk}
We have mentioned that $\langle k^2\rangle/\langle k\rangle$ can be regarded as the measure of network heterogeneity. Hence the threshold in uncorrelated networks is negatively related to the network heterogeneity. According to Theorem~\ref{thm1}, when $2<\gamma\leq 3$, the SF network is heterogeneous enough to make the spreading threshold disappear.
\end{rmk}

Furthermore, in the SF networks, we have the following theorem to characterize the spreading range.
\begin{theorem}
\label{thm2}
In uncorrelated SF networks, the spreading prevalence $\mathcal{P}$ has the following expressions in different cases.
\begin{asparaenum}[(i)]
\item $2<\gamma<3$:
\begin{equation}
\mathcal{P}\sim\rho^{1/(3-\gamma)},\label{eq:3}
\end{equation}
\item $3<\gamma<4$:
\begin{equation}
\mathcal{P}\sim\left(1-\frac{\rho_c}{\rho}\right)^{1/(\gamma-3)},\label{eq:4}
\end{equation}
\item $\gamma>4$ and $\gamma\notin \mathbb{N}$:
\begin{equation}
\mathcal{P}\sim 1-\frac{\rho_c}{\rho},\label{eq:5}
\end{equation}
where ``$\sim$'' means ``be approximately proportional to''. When $\gamma\in\mathbb{N}$, the expression of $\mathcal{P}$ should be discussed case by case. In particular, in the case of $\gamma=3$, we have
\begin{equation}
\mathcal{P}\sim e^{-1/{\rho}}.\label{eq:6}
\end{equation}
\end{asparaenum}
\end{theorem}
\begin{IEEEproof}
See Appendix C.
\end{IEEEproof}

\begin{rmk} We find from Theorem~\ref{thm2} that the spreading prevalence in SF networks is determined by both the model parameters ({i.e.}, $\rho$) and the network topology ({i.e.}, $\gamma$). For example, in an SF network with $2<\gamma<3$, the heterogeneity increases the spreading prevalence if $\rho<1$, while decreases it otherwise. Therefore, the network topology influences the information spreading in complicated manners. We will discuss more details in Section \ref{secthis}.
\end{rmk}

Theorem~\ref{thm1} and Theorem~\ref{thm2} are about the stable state of the spreading process, whereas the following theorem describes its temporal behavior.
\begin{theorem}
\label{thm3}
In uncorrelated networks, the time scale $\tau$ of the spreading process is negatively related to the network heterogeneity, {i.e.}, ${\langle k^2\rangle}/{\langle k\rangle}$, which is the reciprocal of the threshold $\rho_c$. More precisely, we have
\begin{equation}
\tau=\frac{1/\beta}{\rho/\rho_c-1}.\label{eq:7}
\end{equation}
In particular, in the SF networks, the efficiency $\mathcal{E}$ has the following expression:
\begin{equation}
\label{eq:8}
\mathcal{E}=
\begin{cases}
\displaystyle\frac{\alpha\lambda(\gamma-2)-\beta(\gamma-3)}{\gamma-3}&\text{if }\gamma>3,\\
\displaystyle \infty&\text{if }2<\gamma\leq 3.\\
\end{cases}
\end{equation}
\end{theorem}
\begin{IEEEproof}
See Appendix D.
\end{IEEEproof}

\begin{rmk} It is shown in Theorem~3 that an increasing heterogeneity improves the spreading efficiency, namely, it improves the spreading speed. For instance, the infinite efficiency contributes to almost instantaneous rise of the spreading incidence in more heterogeneous SF networks where $2<\gamma\leq3$.
\end{rmk}


\subsection{Correlated Heterogeneous Networks}
\label{secthis}
Previous empirical studies \cite{Mislove2007,Chun2008,Jiang2010} have found that OSNs exhibit assortativity, {i.e.}, vertices with large degrees tend to connect to vertices with large degrees, while vertices with small degrees prefer to connect to those with small degrees. This phenomenon has not been well explained by researchers, and it is conjectured for several reasons, {e.g.}, preferential attachment \cite{Barabasi1999} and proximity bias \cite{Garg2009}. No matter what makes the assortativity in OSNs, it has a significant impact on the edge creation, network evolution and network topology. Thus it is necessary to study how this (positive) second-order correlation affects the dynamic of the information spreading in OSNs. We conduct our analysis on the Markovian networks equipped with both first-order and second-order correlations in this subsection.

We know that $i_k(t)=1-a_k(t)-r_k(t)-q_k(t)$. By omitting\footnote{We cannot perform this approximation all the time as $a_k(t)^2$, $r_k(t)a_k(t)$ and $q_k(t)a_k(t)$ are not always small enough to be omitted during the spreading process. But this approximation works well in this analysis.} terms of $\mathcal{O}(a_k^2)$, Eq.~\ref{eq:0b} can be written as
\begin{equation}
\frac{\mathrm{d}a_k(t)}{\mathrm{d}t}\approx\sum_{k'}L_{kk'}a_{k'}(t),\label{eq:9}
\end{equation}
or
\begin{equation}
\frac{\mathrm{d}\boldsymbol{a}}{\mathrm{d}t}=\boldsymbol{La},\label{eq:10}
\end{equation}
where the Jacobian matrix $\boldsymbol{L}=\{L_{kk'}\}$ is defined by
\begin{equation}
L_{kk'}=-\beta\delta_{kk'}+\alpha\lambda kP(k'\mid k),\label{12}
\end{equation}
and $\delta_{kk'}$ is the Kronecker delta symbol. The solutions of Eq.~\ref{eq:10} imply that the expression of $a_k(t)$ is given by the linear combination of the exponential functions in the form $e^{\lambda_i t}$, where $\lambda_i$ is the eigenvalue of $\boldsymbol{L}$.

Based on the linear stability analysis \cite{Bogua2003} of the system (\ref{eq:10}), we conclude the following theorem which determines the spreading threshold of Markovian networks.
\begin{theorem}
\label{thm4}
The \textbf{connectivity matrix} $\boldsymbol{C}=\{C_{kk'}\}$ is defined by $C_{kk'}=kP(k'\mid k)$. The spreading threshold of Markovian networks is
\begin{equation}
\rho_c=\frac{1}{\Lambda_m},\label{eq:11}
\end{equation}
where $\Lambda_m$ is the largest eigenvalue of $\boldsymbol{C}$.
\end{theorem}
\begin{IEEEproof}
See Appendix E.
\end{IEEEproof}

\begin{rmk} Compared with Theorem~\ref{thm1}, Theorem~\ref{thm4} states a more general result on the spreading threshold in Markovian networks where the degree-degree correlations are considered. In addition, Eq.~\ref{eq:11} agrees with the result of Theorem~\ref{thm1} in the unstructured networks with no second-order correlation, where $\boldsymbol{C}$ has the unique eigenvalue $\Lambda_m=\langle k^2\rangle/\langle k\rangle$.
\end{rmk}

In order to study the influence of the positive degree-degree correlation on the spreading process precisely, here we define the conditional probability $P(k'\mid k)$ satisfying \cite{Vazquez2003}
\begin{equation}
P(k'\mid k)=(1-\theta)q(k')+\theta\delta_{kk'},\label{eq:12}
\end{equation}
where $0\leq\theta<1$. Note that if $\theta=0$, there is no degree-degree correlation in the network. As $\theta$ increases, the network obtains higher positive correlation, {i.e.}, stronger assortativity.

Before stating our main result on the impacts of the degree-degree correlations on the spreading dynamic, we have the following corollary.
\newtheorem{corollary}{\textbf{Corollary}}
\begin{corollary}
\label{cor1}
The spreading threshold of the correlated networks, where degree-degree correlations are defined by Eq.~\ref{eq:12}, is negatively related to the degree-degree correlations, {i.e.},
\begin{equation}
\rho_c\sim1\bigg/\left(\frac{\langle k^2\rangle}{\langle k\rangle}(1-\theta)\right).\label{eq:13}
\end{equation}
\end{corollary}
\begin{IEEEproof}
See Appendix F.
\end{IEEEproof}

\begin{rmk} Corollary~\ref{cor1} implies that higher positive degree-degree correlations increase the spreading threshold. In other words, positive correlations inhibit the information spreading to some extent. Note that when the network is uncorrelated ({i.e.}, $\theta=0$), we also get $\Lambda_m=\langle k^2\rangle/\langle k\rangle$ in the proof of Corollary~\ref{cor1} (see Appendix F).
\end{rmk}

Based on Eq.~\ref{eq:12}, the system of Eqs.~\ref{eq:0a} to \ref{eq:0d} becomes
\begin{IEEEeqnarray}{rCl}
\frac{\mathrm{d} i_k(t)}{\mathrm{d}t}&=&-(1-\theta)\lambda k i_k(t)\sum_{k'}q(k')a_{k'}(t)-\theta\lambda k i_k(t)a_k(t),\IEEEyesnumber\IEEEyessubnumber\label{eq:15a}\\
\frac{\mathrm{d} a_k(t)}{\mathrm{d}t}&=&(1-\theta)\alpha\lambda k i_k(t)\sum_{k'}q(k')a_{k'}(t)+\left(\theta\alpha\lambda k i_k(t)-\beta\right)a_k(t),\IEEEyessubnumber\label{eq:15b}\\
\frac{\mathrm{d} r_k(t)}{\mathrm{d}t}&=&(1-\theta)(1-\alpha)\lambda k i_k(t)\sum_{k'}q(k')a_{k'}(t)+\theta(1-\alpha)\lambda k i_k(t)a_k(t),\IEEEyessubnumber\label{eq:15c}\\
\frac{\mathrm{d} q_k(t)}{\mathrm{d}t}&=&\beta a_k(t),\IEEEyessubnumber\label{eq:15d}
\end{IEEEeqnarray}
In this case, the relations between the network topology and the spreading prevalence are more complicated than those in uncorrelated networks. We have the following theorem.
\begin{theorem}
\label{thm5}
In correlated SF networks, for fixed $\gamma$, we have the following approximation:
\begin{IEEEeqnarray}{rCl}
\mathcal{P}\approx\frac{p_1(\theta)}{p_2(\theta)},\label{eq:a10}
\end{IEEEeqnarray}
where $p_1(\theta)$ and $p_2(\theta)$ are polynomials of $\theta$, and their coefficients are polynomials of $\rho$.
\end{theorem}
\begin{IEEEproof}
See Appendix G.
\end{IEEEproof}

\begin{rmk} Theorem~\ref{thm5} implies that, given $\gamma$, the impacts of $\theta$ on $\mathcal{P}$ are related to $\rho$, {i.e.}, there exists a real number $c_0(\gamma)$ such that the degree-degree correlations promote the spreading prevalence if $\rho<c_0(\gamma)$, whereas inhibit the prevalence if $\rho>c_0(\gamma)$. Note that the opposite case is also possible. Several studies \cite{Moreno2003,Nekovee2007} have concluded incomplete results about the influence of the degree-degree correlations on the epidemic and rumor spreading. In \cite{Nekovee2007}, Nekovee \MakeLowercase{\textit{et al.}} experimentally find that these correlations' impacts on the final fractions of vertices hearing a rumor depend much on the rate of the rumor diffusion. Theorem~\ref{thm5} provides a theoretical explanation behind this phenomenon in the framework of our model. In fact, we have found that the influence of the degree-degree correlations on the information spreading varies a lot along with different information spreading potentials and network degree distributions. 
Note that we do not consider the impact of the largest degree $k_c$ or the network size here. Indeed, $k_c$ and the finite network size affect the spreading process as well \cite{Pastor-Satorras2002}. In Section 6, we will numerically validate that large $k_c$ and the network heterogeneity can balance out the torsion of the degree-degree correlations' influence on spreading caused by the model parameters.
\end{rmk}

We have shown that if $\boldsymbol{L}$ has $n$ distinct eigenvalues, the system $\mathrm{d}\boldsymbol{a}/\mathrm{d}t$ has $e^{\lambda_1 t}\boldsymbol{r}_1$, $\dots$, $e^{\lambda_n t}\boldsymbol{r}_n$ as its fundamental solutions, where $\boldsymbol{r}_i$ is the eigenvector of $\boldsymbol{L}$ with respect to the eigenvalue $\lambda_i$ where $i=1,\dots,n$. Thus, $\boldsymbol{a}(t)$ is determined mainly by the maximum eigenvalue $\lambda_m$, {i.e.},
\begin{equation*}
a_k(t)\sim e^{\lambda_m t}.
\end{equation*}
We have proven part of the following theorem.
\begin{theorem}
\label{thm6}
In correlated networks, the spreading efficiency $\mathcal{E}$ is determined by the maximum eigenvalue of the Jacobian matrix $\boldsymbol{L}$. In the SF networks, we have
\begin{equation}
\mathcal{E}\sim \frac{\langle k^2\rangle}{\langle k\rangle}(1-\theta).\label{eq:16}
\end{equation}
\end{theorem}
\begin{IEEEproof}
See Appendix H.
\end{IEEEproof}

\begin{rmk}
First, in uncorrelated networks, $\boldsymbol{L}$ has the only eigenvalue $\alpha\lambda\langle k^2\rangle/\langle k\rangle-\beta$, thus the conclusion of Theorem~\ref{thm6} agrees with Theorem~\ref{thm3}. Second, the degree-degree correlations inhibit the spreading process in the aspect of the spreading efficiency as well. Meanwhile, the largest degree $k_c$ that contributes to the network heterogeneity also has great influence on the growth time scale. When $k_c\rightarrow\infty$ (usually results in the infinity of $\langle k^2\rangle/\langle k\rangle$), the growth time scale disappears, which implies an irresistible spreading in the correlated networks. Actually, by bounding $\lambda_m$ more precisely, we can get this conclusion without the restriction of specific degree-degree correlations. Given the Perron-Frobenius theorem, we have
\begin{small}
\begin{IEEEeqnarray}{rCl}
\lambda_m^2&\geq&\min_k{\frac{1}{k}\sum_{k'}{k'\sum_l L_{kl}L_{lk'}}}\nonumber\\
&=&\min_k{\sum_l\sum_{k'}\bigg(\alpha^2\lambda^2lP(l\mid k)k'P(k'\mid l)+\beta^2\frac{k'}{k}\delta_{kl}\delta_{lk'}}-\alpha\lambda\beta\frac{l k'}{k}\delta_{kl}P(k'\mid l)-\alpha\lambda\beta k'\delta_{lk'}P(l\mid k)\bigg)\nonumber\\
&=&\min_k\bigg(\sum_l\left(\alpha^2\lambda^2lP(l\mid k)\langle l_{nn}\rangle(k_c)\right)-2\alpha\lambda\beta\langle k_{nn}\rangle(k_c)+\beta^2\bigg)\nonumber\\
&\geq&\min_k \beta^2\left((\rho^2\langle k_{nn}\rangle_{\min}-2\rho)\langle k_{nn}\rangle(k_c)+1\right),\nonumber
\end{IEEEeqnarray}\end{small}where $\langle k_{nn}\rangle_{\min}$ denotes the minimum ANND in the network. Bogu\~n\'a \MakeLowercase{\textit{et al.}} \cite{Bogua2003,Boguna2003} have shown that the function $\langle k_{nn}\rangle(k_c)$ diverges as $k_c\rightarrow\infty$ in the SF networks with $2<\gamma\leq3$, which leads to the divergence of $\lambda_m$ and further the infinite $\mathcal{E}$.
\end{rmk}

\section{Extended Model with Time-Varying Parameters}
\begin{figure}[tb]
\centering
\subfigure[Simulated information spreading based on the naive model]{\includegraphics[width
=3in,height=1.2in]{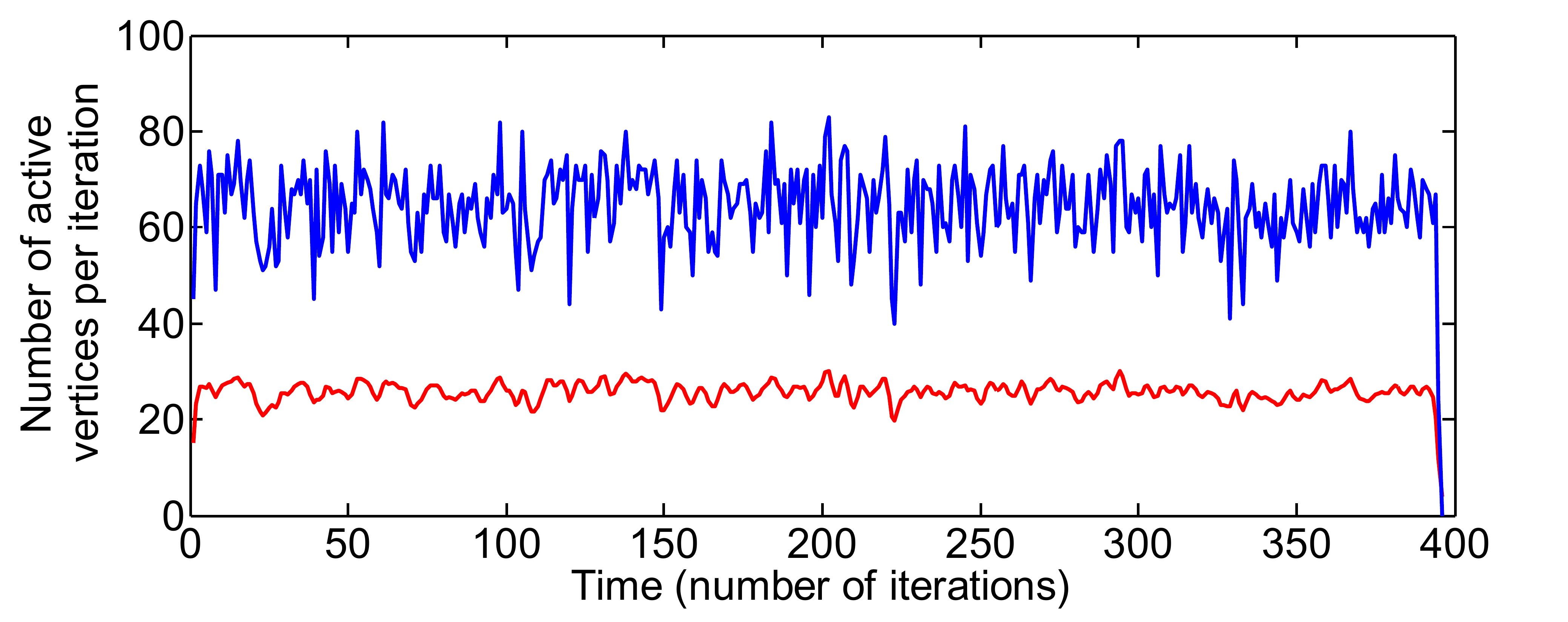}
\label{fig2a}}
\subfigure[A video sharing in Renren]{\includegraphics[width=3in,height=1.2in]{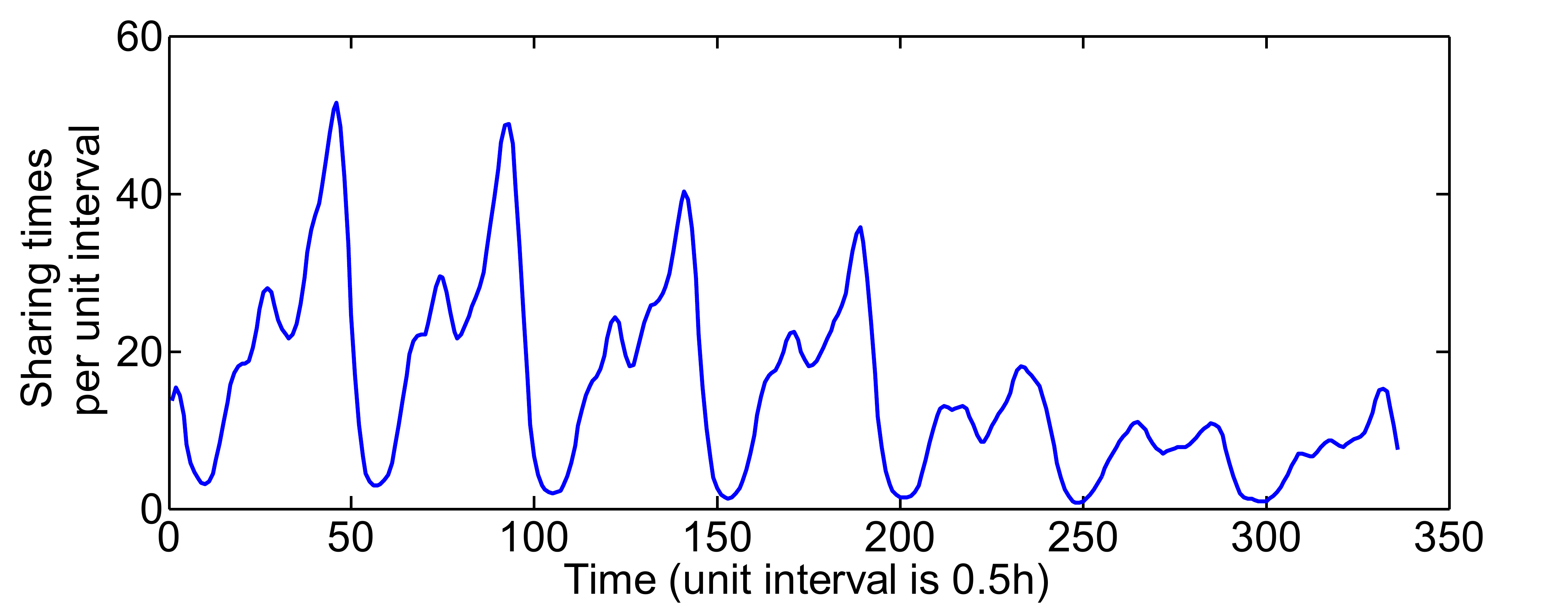}
\label{fig2b}}
\caption{Comparison of the simulated and realistic spreading dynamic. (a) Time series (blue curve) of Monte Carlo simulated information spreading based on the naive model. To reduce rapid fluctuations in the series, we also perform the Gaussian smoothing (red curve). We generate the underlying SF network of 1,000 vertices based on the BA model. We set the model parameters to be $\alpha=0.5$, $\lambda=0.5$ and $\beta=0.5$. (b) Time series of a video sharing in Renren after the Gaussian smoothing.}
\label{fig:2}
\end{figure}

In the naive model, we assume that model parameters $\alpha$, $\lambda$ and $\beta$ are invariant during the spreading process. This assumption is somewhat imprecise considering the realistic circumstances, and we can indeed take one step forward. In fact, there are only three kinds of vertex states in the information spreading process in OSNs, {i.e.}, \emph{ignorant}, \emph{active} and \emph{indifferent}. We add the \emph{quiet} state derived from the \emph{active} state to emphasize and materialize the nonpersistence of the impacts of the active vertices on their neighbors. In addition, it is unreasonable to assume that the probabilities of contacting and propagating the information are invariant from the beginning to the end. In realistic OSNs \cite{Benevenuto2009}, the probability $\lambda$ which describes the likelihood of a user getting the information is related to the temporal variation of user's online behavior, {e.g.}, $\lambda$ varies over time in a day, or even over days of a week. Moreover, as the probability $\alpha$ represents user's interest in the information, it is also varying depending on the fluctuation of the information popularity.

Constant parameters fail to characterize the variation features ({i.e.}, the patterns in the number variation of new coming active vertices over time) in the spreading process. Figure~\ref{fig:2} illustrates the temporal variations of the spreading process in an SF network based on our naive model and of a video sharing in Renren. As shown in Fig.~\ref{fig2b}, we find that in Renren the dynamic of the information spreading owns numerous significant features, {e.g.}, the diurnal pattern, periodicity, sudden spikes and gradual decay, while the time series in Fig.~\ref{fig2a} generated by our naive model seems to be random and lacks notable patterns. This gap between the realistic example and the simulated result leads to our model extension described in the following section.

\subsection{Description of Extended Model}
In our extended spreading model, where $\alpha\triangleq\alpha(t)$ and $\lambda\triangleq\lambda(t)$, we delete the state of \emph{quiet} and the corresponding transition probability $\beta$. Here, the information spreading process terminates along with the vanishing of the information popularity, {i.e.}, $\lim_{t\rightarrow\infty}\alpha(t)=0$.
\begin{figure}
\centering
\scalebox{0.8}{
\begin{tikzpicture}[shorten >=1pt,node distance=1.5cm,auto,semithick]
\node[state] (i)              {$i$};
\node[state] (r) [below right=of i] {$r$};
\node[state] (a) [below left=of i] {$a$};
\path[->]
    (i) edge [loop above] node {$1-\lambda(t)$} (i)
    (i) edge              node {$(1-\alpha(t))\lambda(t)$} (r)
    (i) edge              node [swap] {$\alpha(t)\lambda(t)$} (a);
\end{tikzpicture}
}
\caption{State transition diagram of the extended model. $i$, $a$, $r$ are short for \emph{ignorant}, \emph{active} and \emph{indifferent}, respectively.}
\label{fig:3}
\end{figure}
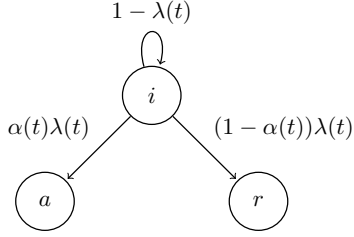

Figure~\ref{fig:3} illustrates the process of the vertex state decay in our extended model. The system Eqs.~\ref{eq:0a} to \ref{eq:0d} can be improved into
\begin{IEEEeqnarray}{rCl}
\frac{\mathrm{d} i_k(t)}{\mathrm{d}t}&=&-\lambda(t) k i_k(t)\sum_{k'}P(k'\mid k)a_{k'}(t),\IEEEyesnumber\IEEEyessubnumber\label{eq:17a}\\
\frac{\mathrm{d} a_k(t)}{\mathrm{d}t}&=&\alpha(t)\lambda(t) k i_k(t)\sum_{k'}P(k'\mid k)a_{k'}(t),\IEEEyessubnumber\label{eq:17b}\\
\frac{\mathrm{d} r_k(t)}{\mathrm{d}t}&=&(1-\alpha(t))\lambda(t) k i_k(t)\sum_{k'}P(k'\mid k)a_{k'}(t),\IEEEyessubnumber\label{eq:17c}
\end{IEEEeqnarray}
where $0\leq\alpha(t)\leq1$, $0\leq\lambda(t)\leq1$ and $\alpha_{\infty}=\lim_{t\rightarrow\infty}\alpha(t)=0$.
\subsection{Impacts of Time-Varying Parameters on Information Spreading}
Since the active vertex is the driving force of the spreading process, here we consider how the time-varying parameters influence the number variation of active vertices. The following theorem characterizes the spreading dynamic.
\begin{theorem}
\label{thm7}
In correlated networks, the derivative of the number of active vertices is given by
\begin{equation}
\frac{\mathrm{d} \boldsymbol{a}(t)}{\mathrm{d}t}=\alpha(t)\lambda(t)\exp\left(\int_0^t{\alpha(x)\lambda(x)\mathrm{d}x}\right)\boldsymbol{C}\boldsymbol{a}(0),\label{eq:18}
\end{equation}
where $\boldsymbol{a}(0)$ is the initial fraction of the active vertices in networks.
\begin{IEEEproof}
See Appendix I.
\end{IEEEproof}
\end{theorem}

\begin{corollary}
In correlated networks, we have
\begin{IEEEeqnarray}{rCl}
\frac{\mathrm{d}a(t)}{\mathrm{d}t}&=&\alpha(t)\lambda(t)\exp\left(\int_0^t{\alpha(x)\lambda(x)\mathrm{d}x}\right)C,\label{eq:21}
\end{IEEEeqnarray}
where $a(t)=\sum_k{P(k)a_k(t)}$, $C=\langle k\rangle\sum_k\sum_{k'}P(k,k')a_{k'}(0)$, which is given by the initial state and the underlying network topology of the information spreading.
\end{corollary}
\begin{IEEEproof}
See Appendix J.
\end{IEEEproof}
\begin{rmk}
Corollary 2 (see Eq.~\ref{eq:21}) implies that the derivative of the number of active vertices, or the spreading velocity, is a function of the time-varying parameters $\alpha(t)$ and $\lambda(t)$ in our extended model.
\end{rmk}
\subsection{Explicit Expressions of $\alpha(t)$ and $\lambda(t)$}
To endow our extended model with more application values, here, we determine the explicit expressions of the time-varying parameters based on several empirical works.

Previous studies \cite{Barabasi2005,Leskovec2007,Chun2008,Cha2009} have shown that the popularity tendency of different kinds of information in OSNs follows a long-tailed or power-law pattern. Probability $\alpha(t)$ quantifies the popularity of one particular information in the following way,
$$\alpha(t)\triangleq \textsf{Pr}\{E_1\text{ happens at time }t\}$$
where $E_1$ represents the event that a user aware of the information becomes active, {i.e.}, the user propagates the information. Note that $E_1$ happens at time $\infty$ means that the user is indifferent, {i.e.}, the user noticing the information does not propagate it. Thus $\alpha(t)$ satisfies $0\leq\alpha(t)\leq1$ and $\int_0^{\infty}\alpha(t)\mathrm{d}t=1$, and is a probability density function. To model the realistic popularity tendency, we let $\alpha(t)$ follow the Gamma distribution $\Gamma(p,\eta)$ whose density function satisfies
\begin{equation}
p(x;p,\eta)=\frac{x^{p-1}e^{-x/\eta}}{\eta^p\Gamma(p)},
\end{equation}
where $x,p,\eta>0$ and $\Gamma(p)$ is the Gamma function. Thus we have $\alpha(t)=p(t;p,\eta)$.

Numerous studies \cite{Gonifmmodemboxccelseccfialves2008,Cha2009,Benevenuto2009} have found that in the Web and SNSes, user's online behaviors are of daily cycle and the diurnal pattern has a gentle peak. We assume that after logging in, the users in SNSes can get all user generated contents (UGCs) which are published and shared by their neighbors immediately. Hence $\lambda(t)$ is determined by the temporal pattern of user's log-in behavior. Similarly, we define $\lambda(t)$ by
$$\lambda(t)\triangleq\textsf{Pr}\{E_2\text{ happens at time }t\},$$
where $E_2$ denotes the event that a user logs in, or equivalently, notice the information propagated by his/her neighbors. We also assume that a user logs in every day and denote $C_p$ as the measure of one day period\footnote{Since the time measure in our model and the real time are incommensurate, here, we set the time period as another parameter.}. Hence $\lambda(t)$ satisfies $0\leq\lambda(t)\leq1$, $\int_0^{C_p}\lambda(t)\mathrm{d}t=1$ and $\lambda(t)=\lambda(t+C_p)$, and is a periodic distribution. We make $\lambda(t)$ follow a popular circular distribution, the \emph{von Mises} distribution, or the \emph{circular normal} \cite{Bishop2006}, whose density distribution is given by
\begin{equation}
p(x;z,\vartheta)=\frac{1}{2\pi I_0(z)}\exp\left(z\cos(x-\vartheta)\right),\label{eq:23}
\end{equation}
where $\vartheta$ corresponds to the mean of the distribution, $z$ is known as the \emph{concentration} parameter and
$$I_0(z)=\frac{1}{2\pi}\int_0^{2\pi}\exp(z\cos\vartheta)\mathrm{d}\vartheta.$$
Note that the von Mises distribution is of period $2\pi$. To model $\lambda(t)$ with period $C_p$, we perform the variable transformation to Eq.~\ref{eq:23} and get
\begin{equation}
\lambda(t)=\frac{1}{C_p I_0(z)}\exp\left(z\cos(\frac{2\pi}{C_p}t-\vartheta)\right).
\end{equation}

In general, user's activity pattern represented by $\lambda(t)$ is fixed on average, and the popularity tendency quantified by $\alpha(t)$ varies for different information. Given the parameter $C$ in Eq.~\ref{eq:21}, we have six unknown parameters to be determined for a specific information spreading process, {i.e.}, $\Theta=\{p,\eta,z,\vartheta,C_p,C\}$. According to Corollary 2, given the initial conditions, we find that the shape of the time series curve $\mathrm{d}a/\mathrm{d}t$ is controlled by $\alpha(t)\lambda(t)\exp\left(\int_0^t{\alpha(x)\lambda(x)\mathrm{d}x}\right)$. 
As this equation is intractable, we adopt the Levenberg-Marquardt (LM) algorithm \cite{Levenberg1944} to determine $\Theta$ (see Section 6.2).

\section{Experiments}
We first present several numerical solutions to the system of Eqs.~\ref{eq:0a} to \ref{eq:0d} based on the naive model proposed in Section 2. These numerical results verify corresponding theoretical results given in Section 4. Furthermore, for our extended model, we use the Levenberg-Marquardt (LM) algorithm \cite{Levenberg1944} to learn key parameters derived in Section 5, based on a dataset of video sharing in Renren. We find that the synthetic evolving curve generated by our extended model matches the realistic time series precisely, which validates the accuracy of our model, and provides an underlying explanation of the dynamic patterns of the information spreading in OSNs. At last, we use Eq.~\ref{eq:21} to predict the spreading tendency of the video sharing given its initial spreading dynamic. Compared with the traditional time series analysis approach, our model works much better and succeeds in predicting future spreading dynamic in the long range.
\subsection{Numerical Results of Naive Model}
We present numerical solutions to the naive model using the standard finite difference scheme on two SF networks of $\gamma=2.5$, $k_c=473$ (heterogeneous) and $\gamma=5$, $k_c=11$ (homogeneous), respectively. The power-law degree sequences of the exponent-varying networks are generated by the algorithm proposed in \cite{Viger2005}. The sizes of these networks are both $n=10,000$. We set the model parameters to be $\lambda=1$ and $\beta=0.3$, and increase the value of $\alpha$ from $0.01$ to $1$ with step-size $0.01$. We investigate how the spreading prevalence $\mathcal{P}$ evolves along with the varying $\alpha$. Note that the value of $\mathcal{P}$ has been averaged over $100$ random equivalent solutions. For each solution, we randomly choose one initial active vertex and leave others ignorant, and meanwhile keep all model and network parameters invariant.
\begin{figure}[tb]
\centering\includegraphics[height=2.9in,width=4.5in]{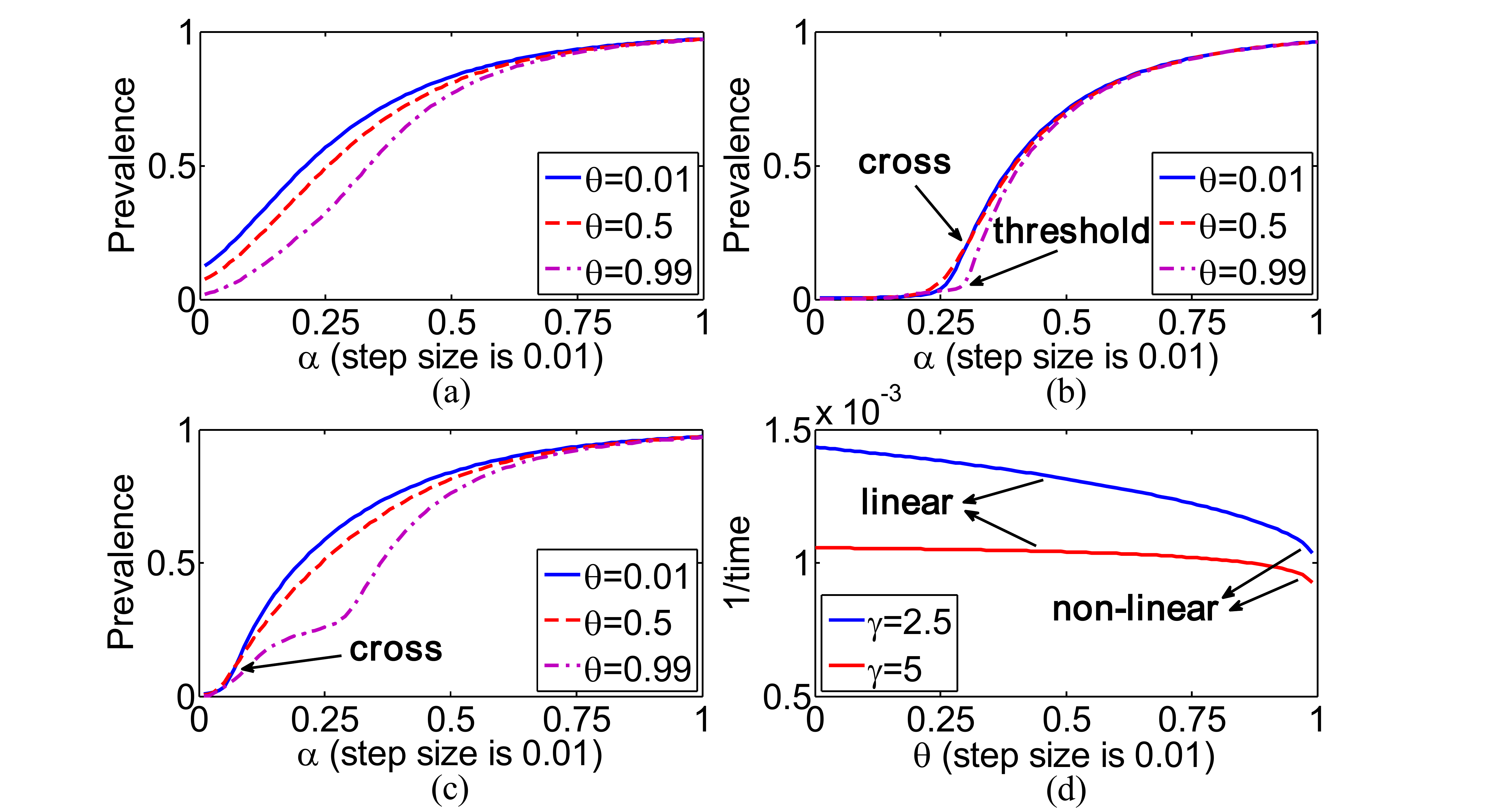}
\caption{Spreading prevalence in correlated SF networks. (a) Prevalence $\mathcal{P}$ in the SF network of $\gamma=2.5$ and $k_c=473$ is plotted as a function of $\alpha$. $\alpha$ varies for every 0.01. The results are shown for three values of the correlation parameter $\theta$. (b) Prevalence $\mathcal{P}$ in the SF network of $\gamma=5$ is plotted as a function of $\alpha$. We find clear turning points and crosses in curves. (c) Prevalence $\mathcal{P}$ in the SF network of $\gamma=2.5$ and $k_c=22$ is plotted as a function of $\alpha$. The crosses reappear in the network with small $k_c$ but the same $\gamma$ as in (a). (d) Reciprocal of the numerical iteration number, which is regarded as the spreading efficiency, is shown for different correlation parameter $\theta$. Two degree distributions of $\gamma=2.5$ and $\gamma=5$ are considered. We find the approximately negative linear relation between the efficiency and degree-degree correlations.}
\label{fig4}
\end{figure}

In Fig.~\ref{fig4}(a), the numerical results for the SF network with $\gamma=2.5$ are shown. As $\lambda$ and $\beta$ have been fixed, the spreading threshold is only related to $\alpha$. In this case, we observe the absence of the threshold, which agrees with the conclusion in Remark~1. Whereas in the case of less heterogeneous network with $\gamma=5$, as shown in Fig.~\ref{fig4}(b), we observe the explicit threshold of the spreading process. Moreover, the threshold is larger in the network with higher positive correlation (i.e., larger $\theta$), which validates the conclusion in Corollary~\ref{cor1} and implies that the positive correlation inhibits the spreading process to some extent.

As for the spreading prevalence, Theorem~\ref{thm5} reveals that the impacts of the correlation on the prevalence relate much to the model parameters and the degree distribution. We find that the curve crosses plotted in Fig.~\ref{fig4}(b) are identified with this conclusion, {i.e.}, \emph{positive correlation increases the spreading prevalence $\mathcal{P}$ in some values of $\alpha$ while decreases $\mathcal{P}$ in other values}. Theorem~\ref{thm5} also holds in the SF network of $\gamma=2.5$, where $k_c$ is small enough\footnote{Generally speaking, the small exponent of the SF network results in the existence of large degree vertices with high probabilities, but they are not negatively related in a deterministic manner.}, {e.g.}, set $k_c=22$ as that in Fig.~\ref{fig4}(c). At the same time, in more heterogeneous networks, {e.g.}, $k_c=473$ in Fig.~\ref{fig4}(a), the positive correlation inhibits the prevalence independently with the model parameters. These complicated relations among the spreading prevalence, model parameters and the network topology ({e.g.}, the largest degree, degree distribution and degree-degree correlations) are partially characterized by Theorem~\ref{thm5}, which as well confirm the essential influence of the largest degree on the information spreading.

In Fig.~\ref{fig4}(d), we set the model parameters to be $\alpha=0.7$, $\lambda=1$ and $\beta=0.3$, and increase the value of $\theta$ from $0$ to $0.99$ with step-size $0.01$. We study how the correlation influences the spreading efficiency. We take the iteration number of solving the system Eqs.~\ref{eq:0a} to \ref{eq:0d} as the spreading time, whose reciprocal is regarded as the spreading efficiency. The results in Fig.~\ref{fig4}(d) are also averaged over $100$ equivalent random solutions. We find approximately negative linear relation between the correlation and the efficiency in both SF networks of $\gamma=2.5$ and $\gamma=5$. Moreover, with the same $\theta$, the information spreads faster in more heterogeneous network (i.e., smaller $\gamma$ and larger $\langle k^2\rangle/\langle k\rangle$). These numerical results conform to the conclusion of Theorem~\ref{thm6}. Note that the non-linear tails of two curves in Fig.~\ref{fig4}(d) may result from the approximation error in the proof of Theorem~\ref{thm6} as $\theta$ becomes large.

\subsection{Learning Dynamic Patterns of Video Sharing in Renren}
From Fig.~\ref{fig:2}, we have seen that the naive model with constant parameters cannot discover the dynamic patterns of information spreading in OSNs. Thus we propose an extended model with time-varying parameters, which can reflect variations in the information popularity and user behavior. In this section, we perform detailed experiments to validate the performance of our extended model in characterizing the dynamic patterns of the realistic information spreading process.

Our video sharing dataset is provided by Renren, one of the most popular SNSes in China. When a user shares a uniform resource locator (URL) of a video from an outside video-sharing site (VSS) to Renren, it becomes a seed and will propagate in the network through being re-shared by the neighbors of the initial sharing user and the neighbors of these neighbors and so on. This process terminates once no one shares the URL any more. We find that this is exactly the process described in Section 2 where our naive model comes from. Note that all kinds of information ({e.g.}, messages, blogs, photos) in Renren propagate in the same manner.

The dataset consists of over 7.5 millions logs on video sharing for a week from September 10th to 16th, 2012. Each log records the sharing time, user identity numbers and the video URLs. And the whole logs contain 335,283 video URLs of which the largest sharing number is 154,955. Note that the number of the videos shared in this period should be smaller than the number of URLs, since the same video may share different URLs. Limited by the dataset, here, we consider each URL as a unique video.

To preprocess the dataset, we first select 1,000 top shared URLs with at least 861 shares. This process is reasonable since the popular URLs propagate in a larger scale than unpopular ones, which reveal more dynamic features in the spreading process. We then construct time series that records the sharing number every thirty minutes based on the selected logs. These time series finally correspond to the curves modelled by Eq.~\ref{eq:21}. In any case, we prefer somewhat smooth curves than the fluctuated ones as we concern more on the general shapes of time series curves, which reflect the essential characteristics of the spreading dynamic. However, the realistic time series oscillate intensely due to countless uncertainties. To reduce these fluctuations and eliminate certain noises in the curve fitting, we apply the Gaussian smoothing on each time series.

Up to now, the experimental verification becomes the problem of curve fitting between the preprocessed time series of the video sharing and the curves determined by Eq.~\ref{eq:21}. It is time-consuming to carry out the curve fitting on every time series and lots of examples share common shapes. This reminds us to first cluster the time series and then fit Eq.~\ref{eq:21} to the centroid of each cluster. Here, we adopte the K-Spectral Centroid (K-SC) clustering algorithm 
proposed by Yang and Leskovec \cite{Yang2011}, which focuses on the pure curve patterns invariant to scaling and translation. In K-SC clustering, the distance $\hat{d}(\boldsymbol{x},\boldsymbol{y})$ between the time series $\boldsymbol{x}$ and $\boldsymbol{y}$ is defined by \cite{Chu1999}
\begin{equation}
\hat{d}(\boldsymbol{x},\boldsymbol{y})=\min_{\nu,h}\frac{\parallel \boldsymbol{x}-\nu \boldsymbol{y}_{(h)}\parallel}{\parallel \boldsymbol{x}\parallel},\label{eq:25}
\end{equation}
where $\boldsymbol{y}_{(h)}$ is the time series after shifting $\boldsymbol{y}$ for $h$ time units and $\parallel\cdot\parallel$ is the $l_2$ norm.
\begin{figure}[tb]
\centering
\subfigure[Average Silhouette]{\includegraphics[width=2.1in,height=1.4in]{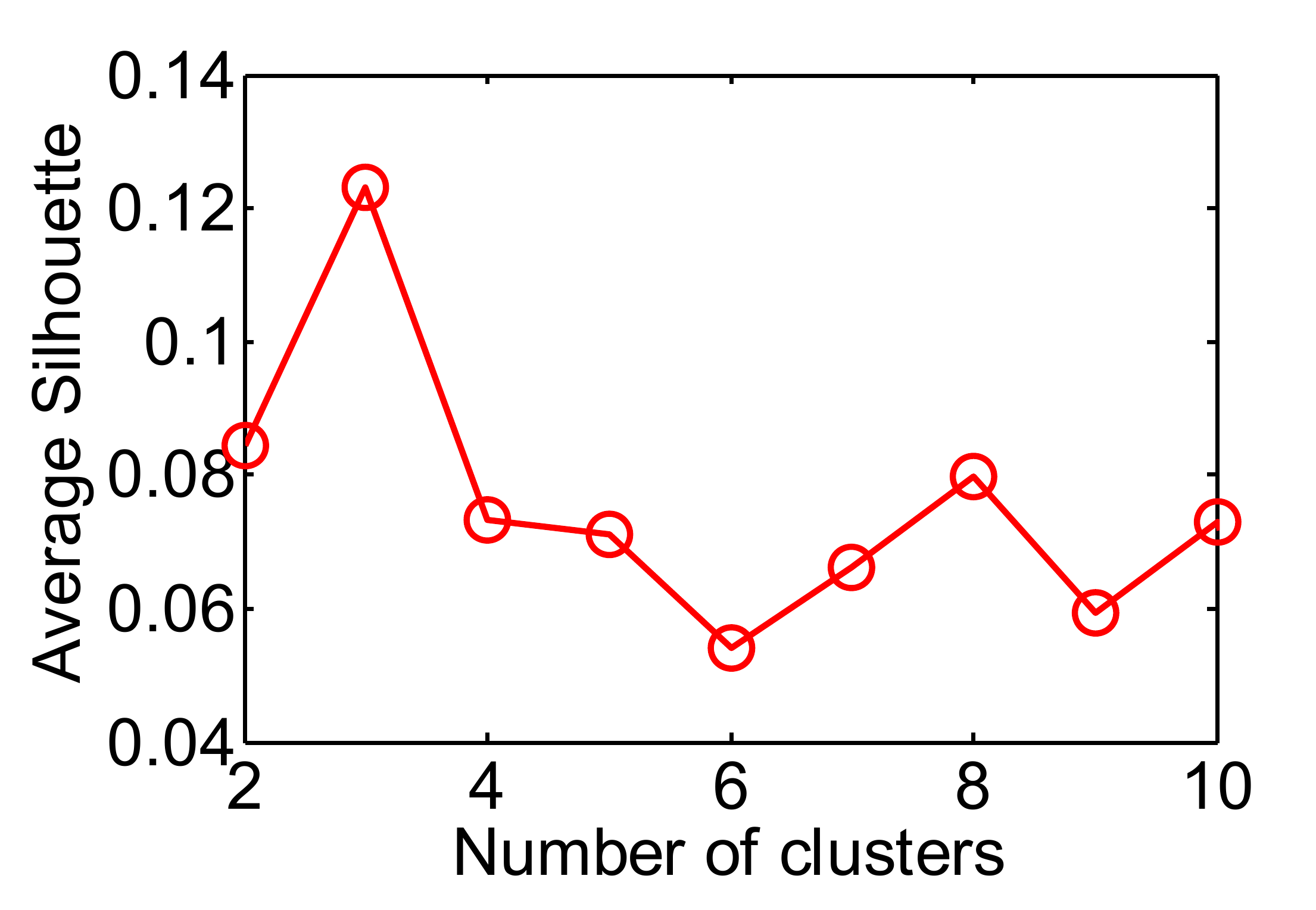}
\label{fig5a}}
\hspace{.15in}
\subfigure[Hartigan¡¯s Index]{\includegraphics[width=2.1in,height=1.4in]{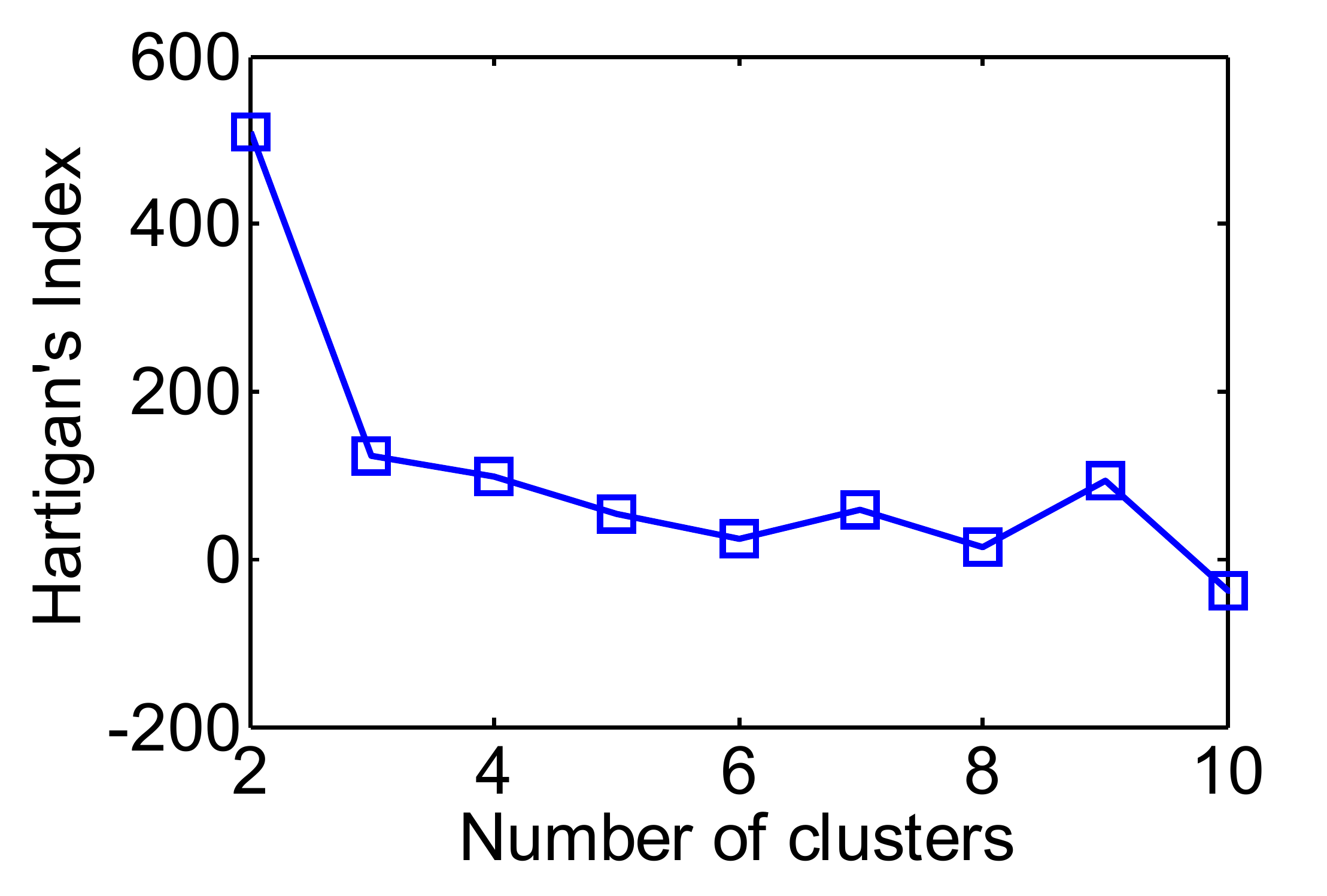}
\label{fig5b}}
\caption{Clustering quality versus the cluster number. (a) The average Silhouette index measures the tightness of the grouped data \emph{in} the cluster and the separability \emph{between} clusters. (b) The Hartigan's index is a relative measure of the square error decrease as the cluster number increases.}
\label{fig5}
\end{figure}
\begin{figure}[tb]
\centering
\subfigure[Centroid of $C_1$]{\includegraphics[width=2.2in,height=0.95in]{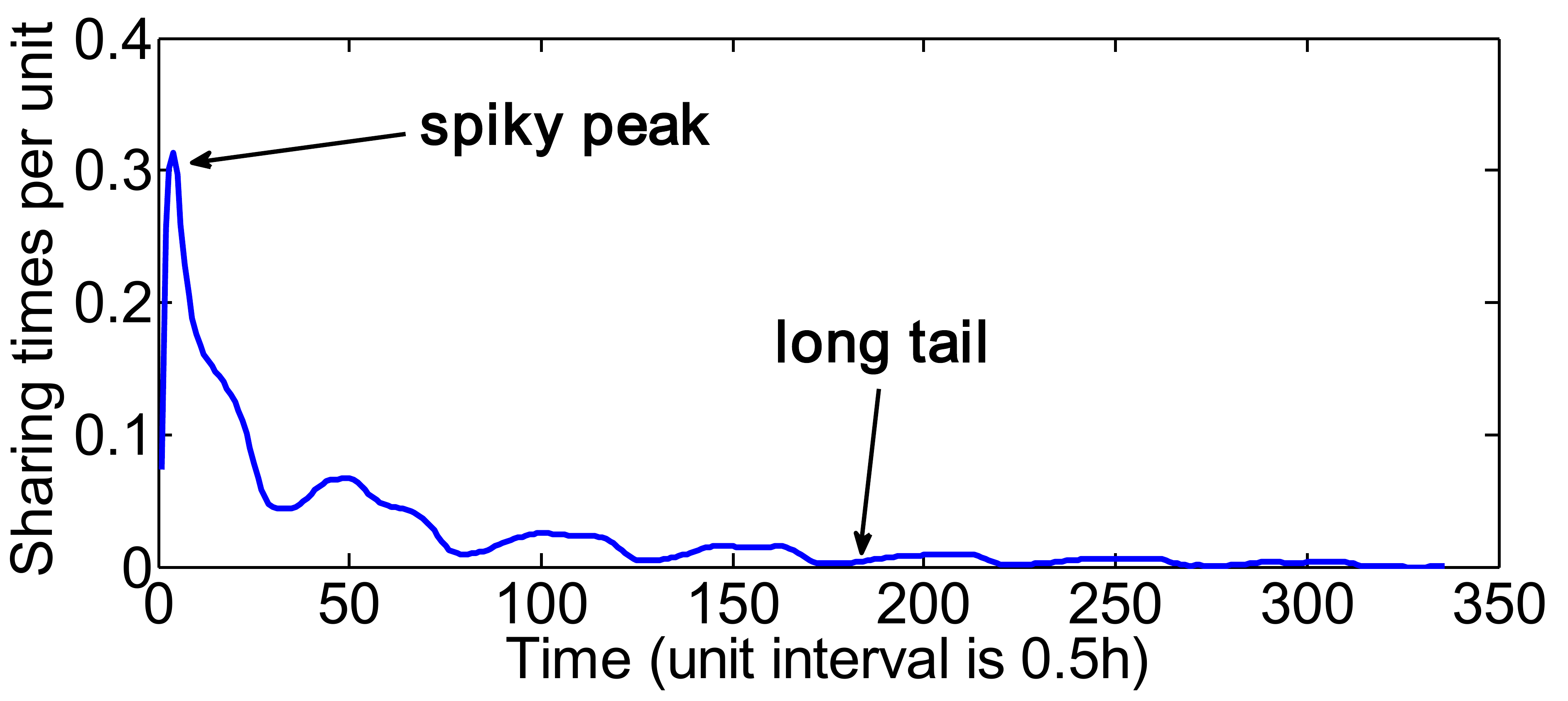}
\label{fig6a}}
\subfigure[Centroid of $C_2$]{\includegraphics[width=2.2in,height=0.95in]{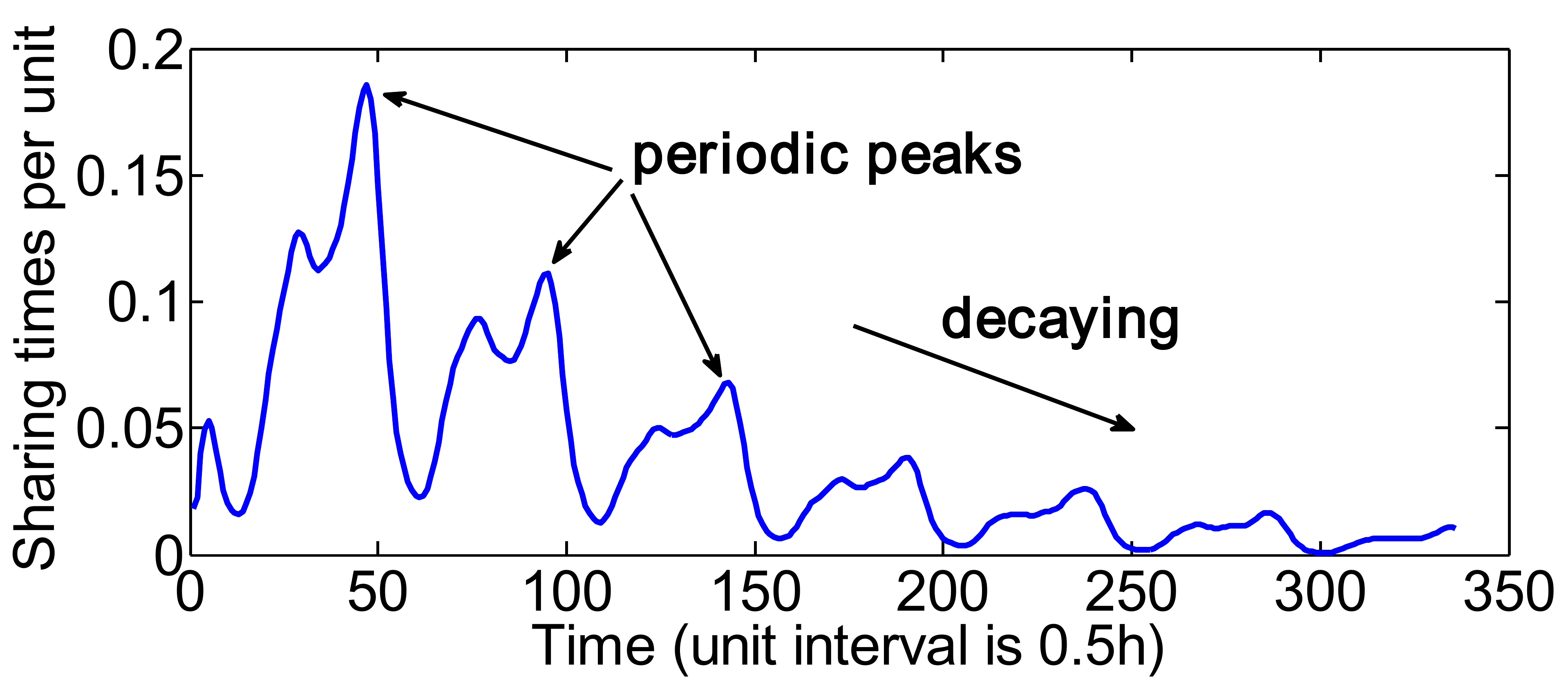}
\label{fig6b}}
\subfigure[Centroid of $C_3$]{\includegraphics[width=2.2in,height=0.95in]{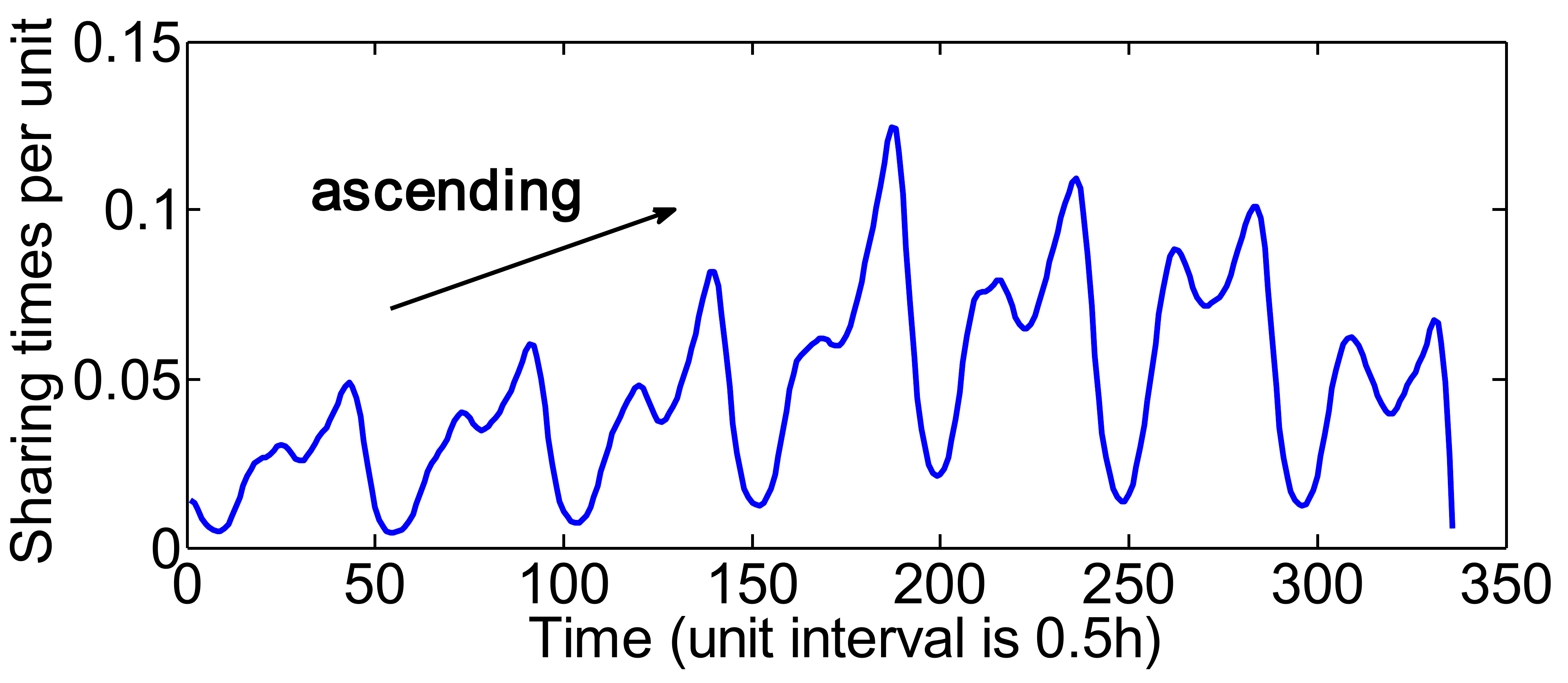}
\label{fig6c}}
\caption{Cluster centroids generated by the K-SC algorithm. The centroids are not realistic time series in the dataset but calculated by K-SC.}
\label{fig6}
\end{figure}

Similar to K-means clustering algorithm, K-SC clustering needs to set the cluster number manually. To obtain the optimal cluster number, we measure the clustering quality in different cluster numbers based on the average Silhouette \cite{Rousseeuw1987} and the Hartigan's index \cite{Hartigan1975}. Note that the distance of the time series in these indices is computed by Eq.~\ref{eq:25}. As shown in Fig.~\ref{fig5}, we find that the two measures on the clustering quality do not agree with each other since they measure the clustering quality in different perspectives and K-SC cannot satisfy both. Hartigan \cite{Hartigan1975} suggests we should choose the smallest number whose Hartigan's index is smaller than a certain value (we set 200 here) as the best cluster number. In addition, the number with the largest average Silhouette is the ideal cluster number. Hence we conclude that \emph{three} is the most reasonable cluster number in our test. The centroids of the clusters $C_1$, $C_2$ and $C_3$ generated after the algorithm converges are presented in Fig.~\ref{fig6}. Note that these centroids are not realistic time series in the dataset, but are computed by the K-SC algorithm and represent the general shapes of the time series curves in each cluster.

We find from Fig.~\ref{fig6} that three centroids express notable dynamic patterns. The time series in cluster $C_1$ own one spiky peak and a long tail which contains several gentle peaks. They may come from the videos that gain users' attentions for a moment and then lose their attractions rapidly. As a limitation, the dataset cannot record the exact whole process for each video sharing, thus some of the time series in $C_1$ may also represent the ending process of the video sharing. The centroid of $C_2$ is similar to the curve in Fig.~\ref{fig2b}, which can be regarded as the typical time series of video sharing. Periodic peaks and gradual decay are general characteristics of dynamic patterns of the video sharing in Renren. Also limited by the dataset, a large number of time series are truncated before terminating, which results in the cluster $C_3$. The time series in $C_3$ could be classified into $C_2$ in a longer period of time. Here, we fit Eq.~\ref{eq:21} to the centroids of $C_1$ and $C_2$ to test our extended model.

\begin{figure}[tb]
\centering
\subfigure[Curve fitting of time series $\boldsymbol{s}_1$ in $C_1$]{\includegraphics[width
=3.0in,height=1.3in]{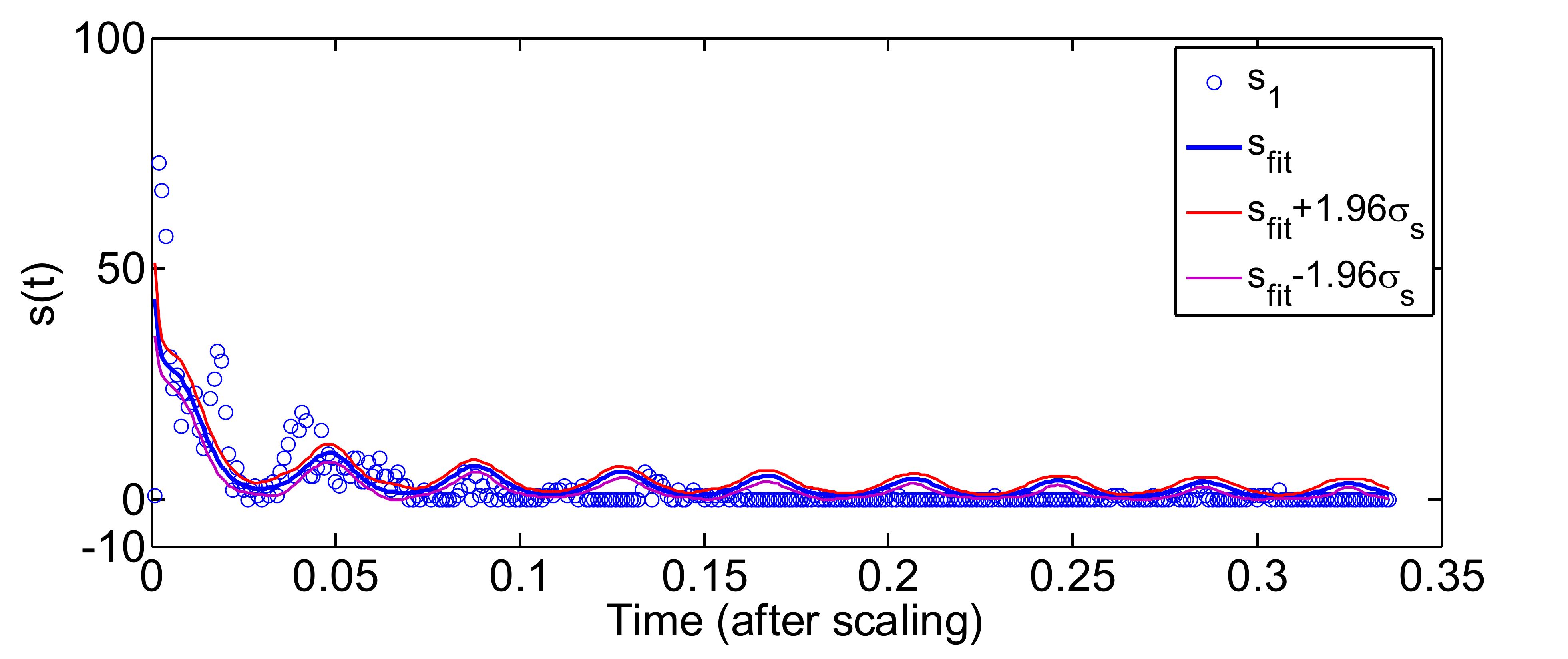}
\label{fig7a}}
\subfigure[Curve fitting of time series $\boldsymbol{s}_2$ in $C_2$]{\includegraphics[width=3.0in,height=1.3in]{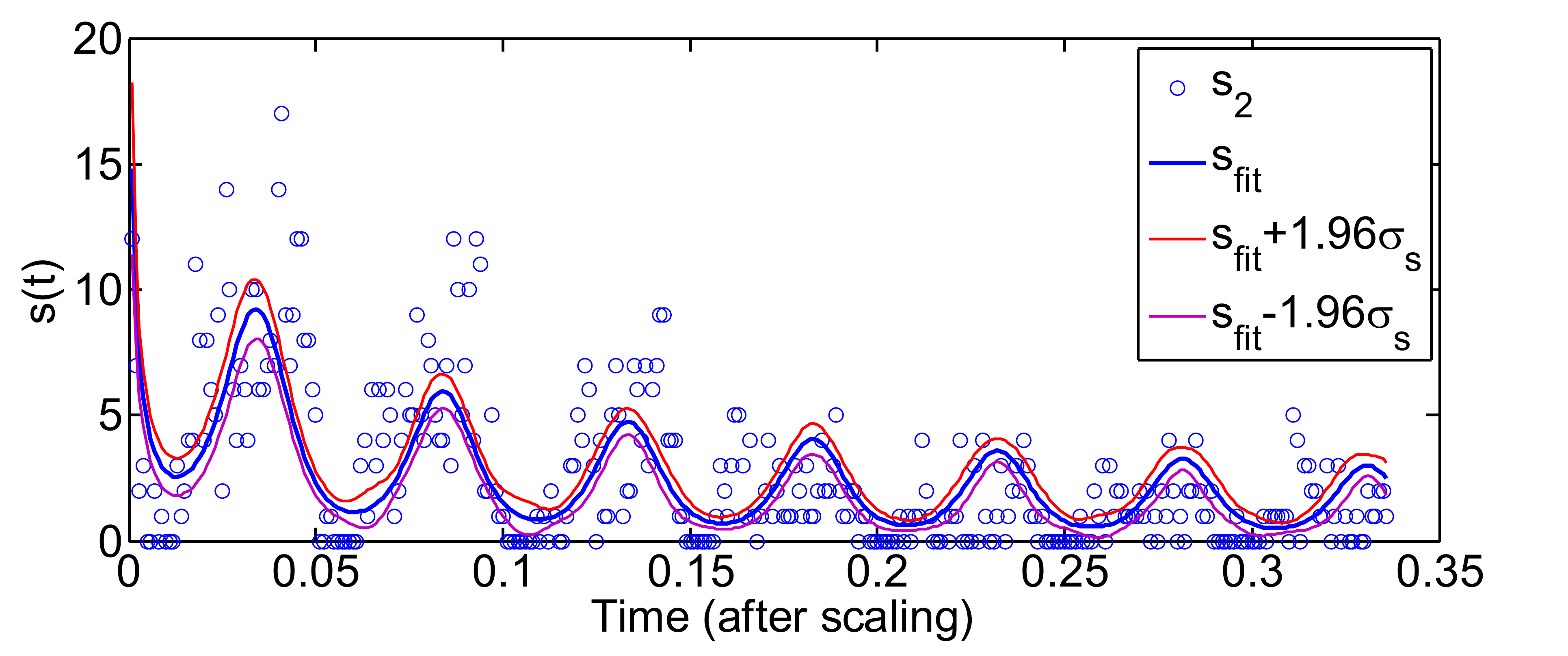}
\label{fig7b}}
\caption{Curve fitting on the representative time series of $C_1$ and $C_2$. The scale between the real time and the model time is adjusted to be 500:1 (in hour). We also plot the 95\% confidence interval (red and purple curves) of the fitting, where $\sigma_{\boldsymbol{s}}$ is the asymptotic standard parameter error that measures how unexplained variability in the data propagates to variability in the parameters.}
\label{fig7}
\end{figure}

We apply the Levenberg-Marquardt (LM) algorithm \cite{Levenberg1944}, a standard algorithm to solve non-linear least squares problems, to perform curve fitting ({i.e.}, to learn the best parameters in Eq.~\ref{eq:21} based on the realistic time series). Because centroids are not realistic time series, we first select the representative time series $\boldsymbol{s}_1$ and $\boldsymbol{s}_2$ of the cluster $C_1$ and $C_2$ by minimizing the distance defined by Eq.~\ref{eq:25}, respectively, {i.e.},
$$\boldsymbol{s}_1=\argmin_{\boldsymbol{s}\in C_1}\hat{d}(\boldsymbol{s},\boldsymbol{c}_1)\text{ and }\boldsymbol{s}_2=\argmin_{\boldsymbol{s}\in C_2}\hat{d}(\boldsymbol{s},\boldsymbol{c}_2),$$
where $\boldsymbol{c}_1$ and $\boldsymbol{c}_2$ are the centroids of $C_1$ and $C_2$, respectively. Note that $\boldsymbol{s}_1$ and $\boldsymbol{s}_2$ are recovered from Gaussian smoothing for fairness. The results of curve fitting on $\boldsymbol{s}_1$ and $\boldsymbol{s}_2$ are presented in Tab.~\ref{table1} and Fig.~\ref{fig7}.
\begin{table}[!t]
\renewcommand{\arraystretch}{1.3}
\caption{Parameter values learned by the LM algorithm}
\label{table1}
\centering
\begin{tabular}{c c c c c c c}
\hline
$\Theta$ & $p$ & $\eta$ & $z$ & $\vartheta$ & $C_p$ & $C$\\
\hline
$\boldsymbol{s}_1$ & 0.4677 & 10.0662 & 0.8443 & 1.4631 & 0.0395 & 0.1586\\
$\boldsymbol{s}_2$ & 0.5157 & 11.5924 & -0.9050 & 1.3159 & 0.0493 & 0.2382\\
\hline
\end{tabular}
\end{table}

As we cannot guarantee the time measure between the model and the realistic scenario are commensurate, we should first calibrate these two time measures. In fact, the spreading process in the model is much faster than that in the real world. 
In our experiments, we set the scale between the real time and the model time to be 500:1 (in hour), which means that one step in the model time amounts to 500 hours in real world. We find from Fig.~\ref{fig7} that Eq.~\ref{eq:21} derived from our extended model fits the time series of video sharing in Renren very well. Specifically, our model can characterize two major features of the information spreading in OSNs, {i.e.}, gradual decay and periodic spiky peaks, via two parameters $\alpha(t)$ and $\lambda(t)$ that depict the information popularity and user behavior, respectively.
Note that limited by somewhat coarseness of the von Mises distribution, Eq.~\ref{eq:21} cannot model more delicate features in the time series, such as two local maxima always appear in one big peak (see Fig.~\ref{fig6b} and Fig.~\ref{fig6c}). We believe that if $\lambda(t)$ is defined by other periodic distributions with two or more local maxima\footnote{However, to our knowledge, there are few periodic distributions of this kind.}, our extended model can characterize more detailed features of the spreading process.\label{sec:6b}
\subsection{Predicting Temporal Dynamic of Video Sharing in Renren}
\begin{figure}[tb]
\centering
\includegraphics[width=3.2in,height=1.45in]{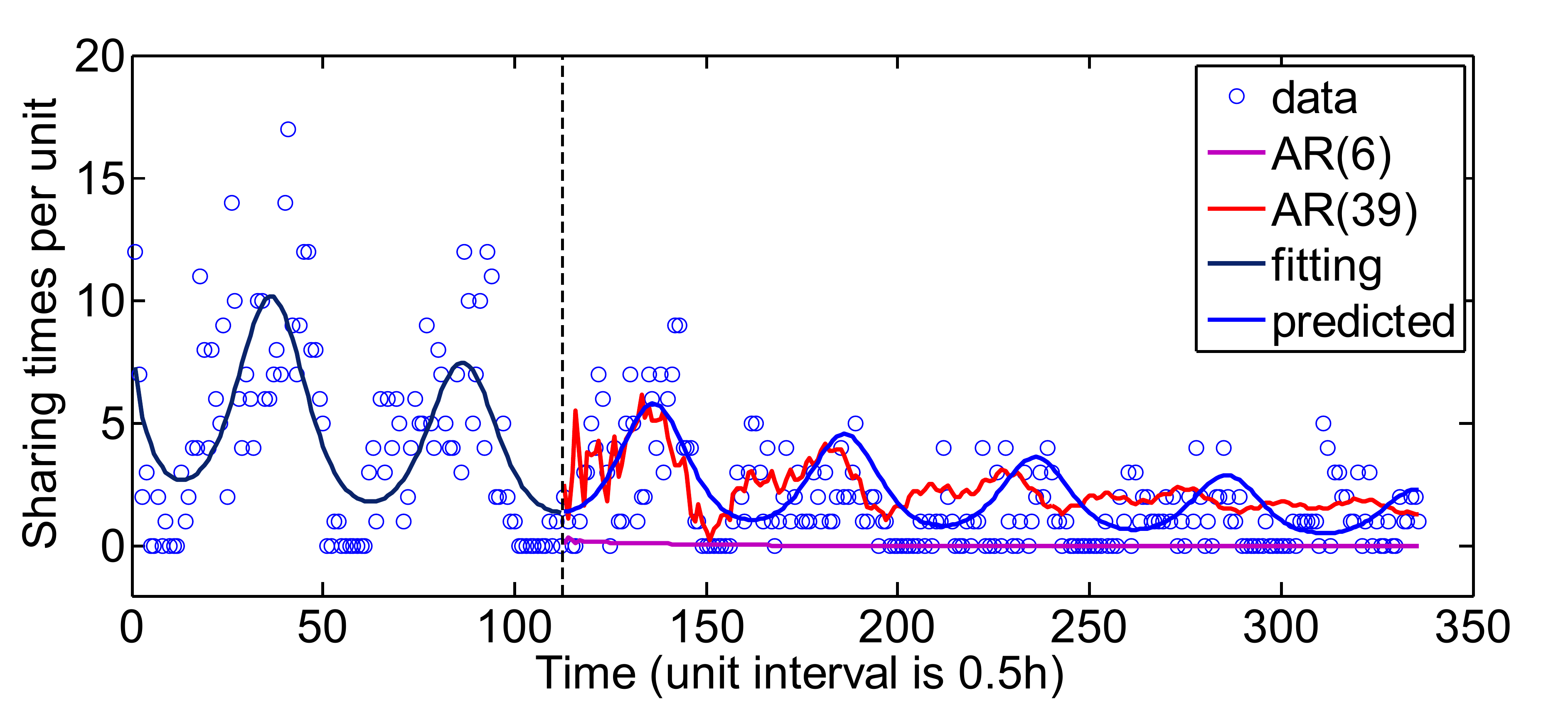}
\caption{Comparison of prediction performance of AR and our extended model on the time series $\boldsymbol{s}_2$. The first 1/3 part (about 56 hours) of $\boldsymbol{s}_2$ is given to train the models. Prediction performance of the remaining 2/3 (about 112 hours) is shown. Note that there are only 6 parameters in our model. For fairness, we set the order ({i.e.}, the number of regression coefficients) of AR as 6 as well. We also increase the order to 39 for a better prediction performance.}
\label{fig:8}
\end{figure}

In addition to characterizing the dynamic patterns of video sharing in Renren, our extended model with time-varying parameters can be used to carry out a more useful task -- to predict future temporal dynamic of the video sharing. Our approach is described as follows: First, given the observation of past time series, we determine the parameters in Eq.~\ref{eq:21} using the LM algorithm. Then Eq.~\ref{eq:21} determines how the time series evolve in the future, which can be used to predict the upcoming part of the time series. We take the representative time series $\boldsymbol{s}_2$ as an example. We first truncate the first 1/3 part of $\boldsymbol{s}_2$ and learn $\Theta$ by the LM curve fitting, which gives $p=0.8822$, $\eta=0.2461$, $z=-0.7737$, $\vartheta=1.5292$, $C_p=0.0498$, $C=0.0586$. Then we use Eq.~\ref{eq:21} equipped with these parameters to generate time series of the remaining 2/3 part. The traditional time series analysis approach, the Autoregressive (AR) model \cite{Box1990}, is used to be compared with our method. Similarly, we estimate the parameters of AR based on the first 1/3 time series and predict the remaining part depending on the trained AR. Detailed results are shown in Fig.~\ref{fig:8}.

From Fig.~\ref{fig:8}, we find that our model is very good at predicting time series in a long range (see the blue curve in Fig.~\ref{fig:8}). While AR with the same number of parameters fails to predict the future spreading tendency and degrades to zero immediately (see the purple curve in Fig.~\ref{fig:8}). In fact, if we increase the order of AR to 39, more than 6 times of the parameter number in our model, AR can characterize the future dynamic as well. But the performance is still not as good as ours. The relative errors \cite{Yang2010}, which are defined by $\sqrt{\sum_t (s(t)-\tilde{s}(t))^2}\big/\sqrt{\sum_t s(t)^2}$ where $\tilde{s}(t)$ is the predicted value, are $0.9828$, $0.7156$, $0.7046$ for AR(6), AR(39) and our model, respectively.
\section{Related Work}
Existing studies on epidemic spreading \cite{Boguna2002,Moreno2002,Pastor-Satorras2002,Boguna2003,Bogua2003,Moreno2003,Barthelemy2004,Barthelemy2005} lay the foundations for studying spreading process in complex networks. Classic epidemic models, such as susceptible-infected-susceptible (SIS) and susceptible-infected-removed (SIR) models, have been studied extensively. Similar to the epidemic spreading, rumor diffusion \cite{DALEY1965,Moreno2004,Nekovee2007,Borge-Holthoefer2013} can be investigated in the same principle. The main difference between the epidemic and rumor spreading is that vertices do not always act actively or react to received messages in the rumor diffusion. This leads to corresponding  modifications for the rumor diffusion models \cite{Borge-Holthoefer2013}. Nevertheless, the mechanisms of the epidemic and rumor spreading are quite different from those of the information spreading in OSNs, which cannot be characterized by these well-studied models.

Many other works model the information spreading process in different assumptions. Two standard models in the innovation diffusion \cite{Rogers2003}, {i.e.}, \emph{Independent Cascades} (IC) and \emph{Linear Threshold} (LT), have been widely employed to model the information spreading in OSNs. Galba \textit{et al.} \cite{Galuba2010} propose a model based on the LT model to predict information cascades in Twitter~\cite{fn7}. Similarly, Guille \textit{et al.} \cite{Guille2012} establish an IC-based model with time-varying parameters to predict the temporal dynamic of information spreading in Twitter. Also for predicting the spreading dynamic, Yang \textit{et al.} \cite{Yang2010} propose the so-called Linear Influence Model (LIM) ignorant with the underlying network topology. As LIM mainly focuses on the spreading dynamic, the influence of vertices is assumed  to be the only factor impacting the spreading process. In addition, Leskovec \textit{et al.} \cite{Leskovec2007} use SIS to model blog citing in the blogosphere, and Cheng \textit{et al.}~\cite{Cheng2014} develop a machine learning framework to predict cascades in OSNs by identifying numerous spreading features. All these models either characterize the information spreading with distinct granularity, or treat the spreading process in different perspectives. However, some of them are too simple to discover the realistic spreading dynamic, and cannot model the dynamic patterns explicitly. Some others do not take the network topology into consideration, which indeed highly influences the information spreading in OSNs. Recently, Gomez \textit{et al.}~\cite{RODRIGUEZ2014} build a probabilistic model to infer the structure and temporal dynamic of the underlying network given information spreading data, which differs from the original motivation of our work. To acquire more related work, readers can refer to \cite{Guille2013}.
%

\section{Conclusion}
In this paper, we first propose a \emph{naive} model based on the IMCs, which features the realistic process of information spreading in OSNs. These IMCs also take the underlying network topology ({e.g.}, degree distribution and degree-degree correlations) into account. Then we derive a dynamical system from the original IMCs based on the mean-field principle. With these coupled differential equations (Eqs.~\ref{eq:0a} to \ref{eq:0d}), we have studied the impacts of the network topology on information spreading in OSNs. 

To discover the temporal dynamic of the information spreading in OSNs, we further propose an \emph{extended} model with time-varying parameters that depict the user behavior and the information popularity. We derive an explicit function (Eq.~\ref{eq:21}) to model the dynamic patterns of the spreading process. Our experiments show that by learning relevant parameters in the function using the curve fitting algorithm, our extended model can capture the dynamic patterns of video sharing in Renren precisely. Given historical measurements, our extended model is also able to predict the long-range spreading tendency accurately, outperforming the standard time series analysis method.


\begin{appendices}
\section{From Probabilistic Model to Dynamical System}
\numberwithin{equation}{section}
\setcounter{equation}{0}

Consider a vertex $n_j$ that is ignorant at time $t$. We denote $p_{ii}^j$ as the probability that $n_j$ stays ignorant in the time interval $[t,t+\Delta t]$. Thus we have $p_{i\tilde{i}}^j=p_{ia}^j+p_{ir}^j=1-p_{ii}^j$, where $p_{i\tilde{i}}^j$, $p_{ia}^j$, $p_{ir}^j$ are probabilities of $n_j$ changing its state, becoming active, and becoming indifferent, respectively. Let $g=g(t)$ denotes the number of active vertices among $n_j$'s neighbors at time $t$, it follows that
\begin{equation*}
p_{ii}^j=\left(1-\Delta t\lambda\right)^g.
\end{equation*}

Assume that the degree of $n_j$ is $k$, and $g$ can be considered as a random variable that has the following binomial distribution,
\begin{equation*}
\Pi(g,t)=\binom{k}{g}\psi(k,t)^g\left(1-\psi(k,t)\right)^{k-g},
\end{equation*}
where $\psi(k,t)$ is the probability at time $t$ that an edge connects an ignorant vertex of degree $k$ with an active vertex. Thus $\psi(k,t)$ can be written as
\begin{equation*}
\psi(k,t)=\sum_{k'}P(k'\mid k)P(a_{k'}\mid i_k)\approx\sum_{k'}P(k'\mid k)a_{k'}(t),
\end{equation*}
where we approximate $P(a_{k'}\mid i_k)$ with $a_{k'}(t)$ by ignoring the correlation between neighboring vertices in different states,

The transition probability $\bar{p}_{ii}(k,t)$ averaged over all possible values of $g$ is given by
\begin{IEEEeqnarray}{rCl}
\bar{p}_{ii}(k,t)&=&\sum_{g=0}^k \binom{k}{g}(1-\Delta t\lambda)^g\psi(k,t)^g(1-\psi(k,t))^{k-g}\nonumber\\
&=&\left(1-\Delta t \lambda\sum_{k'}P(k'\mid k)a_{k'}(t)\right)^k.\label{a1}
\end{IEEEeqnarray}
Based on Eq.~\ref{a1} we get
\begin{IEEEeqnarray}{rCl}
\bar{p}_{i\tilde{i}}(k,t)&=&1-\bar{p}_{ii}(k,t),\label{a2}\\
\bar{p}_{ia}(k,t)&=&\alpha\left(1-\bar{p}_{ii}(k,t)\right),\label{a3}\\
\bar{p}_{ir}(k,t)&=&(1-\alpha)\left(1-\bar{p}_{ii}(k,t)\right)\label{a4}.
\end{IEEEeqnarray}
From Fig.~\ref{fig:1}, we have
\begin{IEEEeqnarray}{rCl}
\bar{p}_{aq}(k,t)&=&\Delta t\beta,\label{a5}\\
\bar{p}_{aa}(k,t)&=&1-\Delta t\beta.\label{a6}
\end{IEEEeqnarray}

Denote $I_k(t)$, $A_k(t)$, $R_k(t)$, $Q_k(t)$ as the expected populations of vertices with degree $k$ that are ignorant, active, indifferent and quiet at time $t$, respectively. The event that an ignorant vertex with degree $k$ becomes active during $[t,t+\Delta t]$ is a Bernoulli random variable with success probability $1-\bar{p}_{ii}(k,t)$. Since the sum of Bernoulli variables follows the binomial distribution with expectation $I_k(t)(1-\bar{p}_{ii}(k,t))$, the difference of the expected population of ignorant vertices with degree $k$ is
\begin{IEEEeqnarray}{rCl}
I_k(t&+&\Delta t)-I_k(t)\nonumber\\
&=&-I_k(t)\left(1-\left(1-\Delta t\lambda\sum_{k'}P(k'\mid k)a_{k'}(t)\right)^k\right).\nonumber\\
\label{a7}
\end{IEEEeqnarray}
Similarly, we can get the difference of populations of active, indifferent and quiet vertices as follows.
\begin{IEEEeqnarray}{rCl}
A_k(t&+&\Delta t)-A_k(t)\nonumber\\
=&\alpha& I_k(t)\left(1-\left(1-\Delta t\lambda\sum_{k'}P(k'\mid k)a_{k'}(t)\right)^k\right)\nonumber\\
&-&\:A_k(t)\Delta t\beta.\label{a8}\\
R_k(t&+&\Delta t)-R_k(t)\nonumber\\
=&(1&-\alpha)I_k(t)\left(1-\left(1-\Delta t\lambda\sum_{k'}P(k'\mid k)a_{k'}(t)\right)^k\right).\nonumber\\
\label{a9}\\
Q_k(t&+&\Delta t)-Q_k(t)=A_k(t)\Delta t\beta.\label{a10}
\end{IEEEeqnarray}

For Eqs.~\ref{a7} to \ref{a10}, we let $\Delta t\rightarrow 0$, omit the higher order infinitesimal and divide each equation by corresponding $N_k$ (the number of vertices with degree $k$), and finally get Eqs.~\ref{eq:0a} to \ref{eq:0d}.
\section{Proof of Theorem 1}
In uncorrelated networks, the conditional probability $P(k'\mid k)$ satisfies
\begin{equation*}
P(k'\mid k)=q(k')=\frac{k'P(k')}{\langle k\rangle}.\label{3.9}
\end{equation*}
Thus the system of Eqs.~\ref{eq:0a} to \ref{eq:0d} degrades into
\begin{IEEEeqnarray}{rCl}
\frac{\mathrm{d} i_k(t)}{\mathrm{d}t}&=&-\lambda k i_k(t)\sum_{k'}q(k')a_{k'}(t),\label{eq:b1}\\
\frac{\mathrm{d} a_k(t)}{\mathrm{d}t}&=&\alpha\lambda k i_k(t)\sum_{k'}q(k')a_{k'}(t)-\beta a_k(t),\label{eq:b2}\\
\frac{\mathrm{d} r_k(t)}{\mathrm{d}t}&=&(1-\alpha)\lambda k i_k(t)\sum_{k'}q(k')a_{k'}(t),\label{eq:b3}\\
\frac{\mathrm{d} q_k(t)}{\mathrm{d}t}&=&\beta a_k(t).\label{3.13}
\end{IEEEeqnarray}

By integrating Eq.~\ref{eq:b1}, we get
\begin{equation}
i_k(t)=i_k(0)e^{-\lambda k\phi(t)},\label{b5}
\end{equation}
where $i_k(0)\approx 1$ and
\begin{equation*}
\phi(t)=\sum_k{q(k)\int_0^t a_k(x)\mathrm{d}x}=\int_0^t\langle\langle a_k(x)\rangle\rangle\mathrm{d}x.
\end{equation*}
Note that we have defined $\langle\langle \star\rangle\rangle\triangleq\sum_k q(k)\star$.

By multiplying Eq.~\ref{eq:b2} with $q(k)$, summing over $k$ and integrating over $t$, we get
\begin{equation}
\frac{\mathrm{d}\phi(t)}{\mathrm{d}t}=\langle\langle a_k(t)\rangle\rangle=\alpha-\alpha\langle\langle e^{-\lambda k\phi(t)}\rangle\rangle-\beta\phi(t).\label{eq:b5}
\end{equation}
Let $t\rightarrow\infty$, we have $\mathrm{d}\phi/\mathrm{d}t\rightarrow0$, and Eq.~\ref{eq:b5} becomes
\begin{equation}
\phi_\infty=\frac{\alpha}{\beta}-\frac{\alpha}{\beta}\sum_k q(k)e^{-\lambda k\phi_\infty},\label{eq:b6}
\end{equation}
where $\phi_\infty=\lim_{t\rightarrow \infty}\phi(t)$.

It is obvious that $\phi_\infty=0 $ is a trivial solution of Eq.~\ref{eq:b6}. The non-zero solution exists if only if the condition
\begin{equation*}
\frac{\mathrm{d}}{\mathrm{d}\phi_\infty}\left(\frac{\alpha}{\beta}-\frac{\alpha}{\beta}\sum_k q(k)e^{-\lambda k\phi_\infty}\right)\Bigg\vert_{\phi_\infty=0}>1
\end{equation*}
is satisfied, i.e.,
\begin{equation}
\frac{\alpha\lambda}{\beta}>\frac{\langle k\rangle}{\langle k^2\rangle}.\label{eq:b7}
\end{equation}
Thus we get the threshold $\rho_c={\langle k\rangle}/{\langle k^2\rangle}$ in uncorrelated networks.

In SF networks, we can calculate $\langle k\rangle$ and $\langle k^2\rangle$ as follows.
\begin{IEEEeqnarray}{rCl}
\langle k\rangle&=&\sum_{k=1}^{\infty}kP(k)=Z\sum_{k=1}^{\infty}k^{-\gamma+1}\nonumber\\
&\approx& Z\int_1^{\infty}k^{-\gamma+1}\mathrm{d}k=\frac{\gamma-1}{2-\gamma} k^{-\gamma+2}\bigg\vert_1^{\infty},\label{eq:b8}
\end{IEEEeqnarray}
\begin{IEEEeqnarray}{rCl}
\langle k^2\rangle&=&\sum_{k=1}^{\infty}k^2P(k)=Z\sum_{k=1}^{\infty}k^{-\gamma+2}\nonumber\\
&\approx& Z\int_1^{\infty}k^{-\gamma+2}\mathrm{d}k=\frac{\gamma-1}{3-\gamma} k^{-\gamma+3}\bigg\vert_1^{\infty}.\label{eq:b9}
\end{IEEEeqnarray}

To get the reasonable result, the networks we consider should have finite average degree. Thus according to Eq.~\ref{eq:b8}, we have $\gamma>2$. In addition, $\langle k^2\rangle<\infty$ if $\gamma>3$, and $\langle k^2\rangle=\infty$ otherwise. Hence the spreading threshold of uncorrelated SF networks is given by Eq.~\ref{eq:2}.
\section{Proof of Theorem 2}
We consider uncorrelated SF networks in this theorem. Denote the spreading prevalence of each degree $k$ as $\mathcal{P}_k=r_k(\infty)+q_k(\infty)$, and we have $\mathcal{P}=\sum_k P(k)\mathcal{P}_k$. Given Eq.~\ref{b5}, we have
\begin{IEEEeqnarray}{rCl}
\mathcal{P}&=&\sum_k{Zk^{-\gamma}\left(1-e^{-\lambda k\phi_{\infty}}\right)}\nonumber\\
&\approx&1-(\gamma-1)\int_1^{\infty}{k^{-\gamma}e^{-\lambda k\phi_{\infty}}\mathrm{d}k}\nonumber\\
&=&1-(\gamma-1)z^{\gamma-1}\int_z^{\infty}x^{-\gamma}e^{-x}\mathrm{d}x\nonumber\\
&=&1-(\gamma-1)z^{\gamma-1}\Gamma(-\gamma+1,z),\label{eq:c1}
\end{IEEEeqnarray}
where $z=\lambda \phi_{\infty}$, $x=\lambda k\phi_{\infty}$ and $\Gamma(s,z)$ is the \emph{incomplete Gamma function}.

As $\Gamma(s,z)$ can be written as
\begin{equation}
\Gamma(s,z)=\Gamma(s)-\int_0^z{x^{s-1}e^{-x}\mathrm{d}x},\label{eq:c2}
\end{equation}
we perform the Taylor expansion on the integrand of the right-hand side of Eq.~\ref{eq:c2} for small $z$ and get
\begin{equation}
\Gamma(s,z)=\Gamma(s)-\frac{z^s}{s}-z^s\sum_{n=1}^{\infty}{\frac{(-z)^n}{(s+n)n!}},\label{eq:c3}
\end{equation}
where $\Gamma(s)$ is the \emph{standard Gamma function}. Note that this expansion makes sense only if $s\neq 0,-1,-2,\dots$. The cases of $\gamma\in\mathbb{N}$ should be discussed case by case.

By substituting Eq.~\ref{eq:c3} into Eq.~\ref{eq:c1}, we get
\begin{IEEEeqnarray}{rCl}
\mathcal{P}&=&\Gamma(-\gamma+2)z^{\gamma-1}+(\gamma-1)\sum_{n=1}^{\infty}\frac{(-z)^n}{(n-\gamma+1)n!}\nonumber\\
&\approx& \frac{\gamma-1}{\gamma-2}z+\mathcal{O}(z^{\gamma-1}).\nonumber
\end{IEEEeqnarray}
Thus for any $\gamma>2$, we have
\begin{equation}
\mathcal{P}\approx\frac{\gamma-1}{\gamma-2}z=\frac{\gamma-1}{\gamma-2}\lambda \phi_{\infty}.\label{eq:c4}
\end{equation}

We will get the expression of $\mathcal{P}$ once $\phi_{\infty}$ is obtained. For Eq.~\ref{eq:b6}, we perform the same calculations as Eq.~\ref{eq:c1} and get
\begin{equation*}
\phi_{\infty}=\frac{\alpha}{\beta}-\frac{\alpha}{\beta}(\gamma-2)z^{\gamma-2}\Gamma(-\gamma+2,z),
\end{equation*}
which gives
\begin{equation}
\phi_{\infty}=\frac{\alpha}{\beta}z^{\gamma-2}\Gamma(3-\gamma)+\frac{\alpha}{\beta}(\gamma-2)\sum_{n=1}^{\infty}\frac{(-z)^n}{(n+2-\gamma)n!}.\label{eq:c5}
\end{equation}
The leading behavior of $\phi_{\infty}$ depends on particular values of $\gamma$. We consider the following cases.

(i) $2<\gamma<3$:

In this case, we have
\begin{equation*}
\phi_\infty\approx\frac{\alpha}{\beta}(\lambda \phi_{\infty})^{\gamma-2}\Gamma(3-\gamma),
\end{equation*}
which gives
\begin{equation}
\phi_\infty\approx\left(\Gamma(3-\gamma)\right)^{1/(3-\gamma)}\left(\alpha/\beta\right)^{1/(3-\gamma)}\lambda^{(\gamma-2)/(3-\gamma)}.\label{eq:c6}
\end{equation}
Combining Eq.~\ref{eq:c6} with Eq.~\ref{eq:c4}, we get Eq.~\ref{eq:3}.

(ii) $3<\gamma<4$:

According to Eq.~\ref{eq:c5}, we have
\begin{equation*}
\phi_{\infty}\approx\frac{\alpha}{\beta}(\lambda \phi_{\infty})^{\gamma-2}\Gamma(3-\gamma)-\frac{\alpha}{\beta}\frac{\gamma-2}{3-\gamma} \lambda\phi_{\infty},
\end{equation*}
which gives
\begin{equation}
\phi_{\infty}\approx\left(\frac{\beta}{\alpha\lambda^{\gamma-2}}\frac{\gamma-2}{\Gamma(4-\gamma)}\left(\frac{\alpha\lambda}{\beta}-\frac{\gamma-3}{\gamma-2}\right)\right)^{1/(\gamma-3)}.\label{eq:c7}
\end{equation}
Combining Eq.~\ref{eq:c7} with Eq.~\ref{eq:c4}, we get Eq.~\ref{eq:4}.

(iii) $\gamma>4$:

The dominant terms in the expansion of $\phi_{\infty}$ now becomes
\begin{equation*}
\phi_\infty\approx\frac{\alpha}{\beta}\frac{\gamma-2}{\gamma-3}\lambda\phi_{\infty}-\frac{\alpha}{2\beta}\frac{\gamma-2}{\gamma-4}(\lambda \phi_{\infty})^2,
\end{equation*}
which gives
\begin{equation}
\phi_{\infty}\approx\frac{2(\gamma-4)}{\gamma-3}\frac{\beta}{\alpha\lambda^2}\left(\frac{\alpha\lambda}{\beta}-\frac{\gamma-3}{\gamma-2}\right).\label{eq:c8}
\end{equation}
Combining Eq.~\ref{eq:c8} with Eq.~\ref{eq:c4}, we get Eq.~\ref{eq:5}.

(iv) $\gamma=3$:

By carrying out similar approximation techniques as those in cases (i) to (iii) except making use of the expansion of $\Gamma(0,z)$ as $z\rightarrow0$, {i.e.},
\begin{equation}
\Gamma(0,z)=-(\gamma_E+\ln(z))+z+\mathcal{O}(z^2),\label{eq:c9}
\end{equation}
where $\gamma_E$ is the Euler's constant, instead of the expansion Eq.~\ref{eq:c3}, we can get Eq.~\ref{eq:6}. For brevity, we omit the detailed proof here.
\section{Proof of Theorem 3}
By omitting terms of $\mathcal{O}(a^2)$, we can write Eq.~\ref{eq:b2} as
\begin{IEEEeqnarray}{rCl}
\frac{\mathrm{d} a_k(t)}{\mathrm{d}t}&\approx&\alpha\lambda k \sum_{k'}q(k')a_{k'}(t)-\beta a_k(t)\nonumber\\
&=&\alpha\lambda k \langle\langle a_{k}(t)\rangle\rangle-\beta a_k(t),\label{eq:d1}
\end{IEEEeqnarray}
where $\langle\langle a_{k}(t)\rangle\rangle$ is a function of $t$ with $\langle\langle a_{k}(0)\rangle\rangle\approx a(0)\approx1$. Then the derivative of $\langle\langle a_k(t)\rangle\rangle$ can be written as
\begin{IEEEeqnarray}{rCl}
\frac{\mathrm{d}\langle\langle a_k(t)\rangle\rangle}{\mathrm{d}t}&=&\sum_k q(k)\frac{\mathrm{d}a_k(t)}{\mathrm{d}t}\nonumber\\
&=&\sum_k q(k)\left(\alpha\lambda k \langle\langle a_k(t)\rangle\rangle-\beta a_k(t)\right)\nonumber\\
&=&\alpha\lambda\left(\frac{\langle k^2\rangle}{\langle k\rangle}-\frac{\beta}{\alpha\lambda}\right)\langle\langle a_k(t)\rangle\rangle.\label{eq:d2}
\end{IEEEeqnarray}

By solving Eqs.~\ref{eq:d2} and \ref{eq:d1}, we have
\begin{equation}
\langle\langle a_k(t)\rangle\rangle=e^{t/\tau}
\end{equation}
and
\begin{equation}
a_k(t)=e^{-\beta t}\left(1+\frac{k\langle k\rangle}{\langle k^2\rangle}\left(e^{\left(\frac{1}{\tau}+\beta\right)t}-1\right)\right),\label{eq:d4}
\end{equation}
where $\tau$ is given by Eq.~\ref{eq:7}.

For the number of active vertices $a(t)$ which equals to $\sum_k P(k)a_k(t)$, we have
\begin{equation}
a(t)=e^{-\beta t}\left(1+\frac{{\langle k\rangle}^2}{\langle k^2\rangle}\left(e^{\left(\frac{1}{\tau}+\beta\right)t}-1\right)\right).\label{eq:d5}
\end{equation}

Note that $e^{{t}/{\tau}}$ is the dominant term of Eqs.~\ref{eq:d4} and \ref{eq:d5}. Furthermore, for uncorrelated SF networks, Eq.~\ref{eq:8} can be get easily from Eq.~\ref{eq:7} given Eqs.~\ref{eq:b8} and \ref{eq:b9}. This completes the proof.
\section{Proof of Theorem 4}
We consider correlated networks (i.e., Markovian networks) in this theorem. Consider the Jacobian matrix $\boldsymbol{L}$ defined by Eq.~\ref{12}, the solution $\boldsymbol{a}=\boldsymbol{0}$ is unstable if there exists at least one positive eigenvalue of $\boldsymbol{L}$.

Based on Eq.~\ref{3}, we have the connectivity detailed balance condition \cite{Boguna2002}
\begin{equation}
kP(k'|k)P(k)=k'P(k|k')P(k')=\langle k\rangle P(k,k').\label{eq:e1}
\end{equation}
As to the connectivity matrix $\boldsymbol{C}$, if $(v_k)$ is an eigenvector of $\boldsymbol{C}$ with eigenvalue $\Lambda$, then by Eq.~\ref{eq:e1}, $(P(k)v_k)$ is an eigenvector of $\boldsymbol{C}^{\textsf{T}}$ with the same eigenvalue. Hence all eigenvalues of $\boldsymbol{C}$ are real. Let $\Lambda_m$ be the largest eigenvalue of $\boldsymbol{C}$. Since $\boldsymbol{L}=\alpha\lambda\boldsymbol{C}-\beta\boldsymbol{I}$, the largest eigenvalue of $\boldsymbol{L}$ is $\alpha\lambda\Lambda_m-\beta$. Therefore, $\boldsymbol{L}$ has at least one positive eigenvalue whenever $\alpha\lambda\Lambda_m-\beta>0$. This gives the threshold $\rho_c=1/\Lambda_m$.

\section{Proof of Corollary 1}
The connection matrix $\boldsymbol{C}$ of SF networks can be written as $\boldsymbol{C}=\left\{(1-\theta)kq(k')+\theta k\delta_{kk'}\right\}$. Thus we have
\begin{equation*}
\det(\boldsymbol{C}-\Lambda \boldsymbol{I})=f(\Lambda)\cdot\prod_{k=2}^{k_c}(\theta k-\Lambda),
\end{equation*}
where
\begin{equation*}
f(\Lambda)=(\theta-\Lambda)\left(1+(1-\theta)\sum_{k=1}^{k_c}\frac{kq(k)}{\theta k-\Lambda}\right).
\end{equation*}

With regard to $f(\Lambda)$, we find that for any $\theta$ and $k=1,\dots,k_c$, there is null point $\Lambda_k$ locating in $(\theta k,\theta k+\varepsilon_k)$, where $\theta k$ is the pole of $f(\Lambda)$ and $\varepsilon_k>0$. Denote $\Lambda_{m}\in(\theta k_c,\theta k_c+\varepsilon_{k_c})$ as the maximum null point of $f(\Lambda)$, hence it is the largest eigenvalue of $\boldsymbol{C}$. To approximate $\Lambda_{m}$, we have
$$1+(1-\theta)\sum_{k=1}^{k_c}\frac{kq(k)}{\theta k-\Lambda_m}=0,$$
which further yields
$$\Lambda_m-(1-\theta)\sum_{k=1}^{k_c}\frac{kq(k)}{1-\frac{\theta k}{\Lambda_m}}\approx\Lambda_m-\frac{\langle k^2\rangle}{\langle k\rangle}(1-\theta)=0.$$
This implies Eq.~\ref{eq:13}.

\section{Proof of Theorem 5}
The proof techniques used here are similar to those in the proof of Theorem~\ref{thm2}, see Appendix C.

\numberwithin{equation}{section}
\setcounter{equation}{0}
By integrating Eq.~\ref{eq:15a}, we get
\begin{equation}
i_k(t)=i_k(0)e^{-\lambda k\phi_k(t)},\label{eq:a1}
\end{equation}
where
\begin{IEEEeqnarray}{rCl}
\phi_k(t)&=&(1-\theta)\phi(t)+\theta\int_0^t a_k(x)\mathrm{d}x\label{eq:a2}
\end{IEEEeqnarray}
and $\phi(t)$ is defined in Appendix B.
Given $a_k(0)\approx 0$, we integrate Eq.~\ref{eq:15b} and have
\begin{equation}
a_k(t)=\alpha-\alpha i_k(t)-\beta\int_0^t a_k(x)\mathrm{d}x.\nonumber
\end{equation}
Since $a_k(\infty)=a_\infty=0$, we have
\begin{equation}
\int_0^{\infty} a_k(x)\mathrm{d}x=\frac{\alpha}{\beta}\left(1-i_k(\infty)\right).\label{eq:a4}
\end{equation}
By combining Eqs.~\ref{eq:a1}, \ref{eq:a2} and \ref{eq:a4}, we have
\begin{IEEEeqnarray}{rCl}
\phi_k(\infty)&=&(1-\theta)\phi_{\infty}+\theta\int_0^{\infty} a_k(x)\mathrm{d}x\nonumber\\
&=&(1-\theta)\phi_{\infty}+\frac{\alpha\theta}{\beta}\left(1-e^{-\lambda k\phi_k(\infty)}\right).\label{eq:a5}
\end{IEEEeqnarray}
By performing the Taylor expansion on $e^{-\lambda k\phi_k(\infty)}$ of Eq.~\ref{eq:a5} and omitting
 higher-order terms, we get
\begin{equation}
\phi_{k}(\infty)=\frac{1-\theta}{1-\theta\rho k}\phi_{\infty}.\label{eq:a6}
\end{equation}

Given Eqs.~\ref{eq:a1} and \ref{eq:a6}, we have
\begin{IEEEeqnarray}{rCl}
\mathcal{P}&=&\sum_k{P(k)\mathcal{P}_k}=\sum_k{Zk^{-\gamma}\left(1-e^{-\lambda k\phi_k(\infty)}\right)}\nonumber\\
&=&1-(\gamma-1)\int_1^{\infty}{k^{-\gamma}e^{-\frac{(1-\theta)\lambda k}{1-\theta\rho k}\phi_{\infty}}\mathrm{d}k},\label{eq:a7}
\end{IEEEeqnarray}
where $1/(1-\theta\rho k)$ can be expanded as
\begin{equation*}
\frac{1}{1-\theta\rho k}=\sum_{n=0}^{\infty}{\left(\theta\rho k\right)^n}
\end{equation*}
when $\theta\rho k<1$. At the same time, part of the integrand in Eq.~\ref{eq:a7}, $\exp\left(-\frac{(1-\theta)\lambda k}{1-\theta\rho k}\phi_{\infty}\right)$, can be written as
\begin{equation*}
e^{-\frac{(1-\theta)\lambda k}{1-\theta\rho k}\phi_{\infty}}=e^{\left(\frac{(1-\theta)\lambda}{\theta\rho}\phi_{\infty}\right)\big/\left(1-\frac{1}{\theta\rho k}\right)},
\end{equation*}
where $1\Big/\left(1-\frac{1}{\theta\rho k}\right)$ has the expansion
\begin{equation*}
1\bigg/\left(1-\frac{1}{\theta\rho k}\right)=\sum_{n=0}^{\infty}{\left(\frac{1}{\theta\rho k}\right)^n}
\end{equation*}
when $\theta\rho k>1$.

Now we analyze relations between $\mathcal{P}$ and $\theta$ on a case-by-case basis.

(i) $\gamma>2$ and $\gamma\notin\mathbb{N}$:

In this case, we have the following Eq.~\ref{eq:a8},
\begin{IEEEeqnarray}{rCl}
\mathcal{P}&=&1-(\gamma-1)\left(\int_1^{\lfloor k_1\rfloor}{{k^{-\gamma}e^{- (1-\theta)\lambda k\phi_{\infty}\sum_{n=0}^{\infty}{\left(\theta\rho k\right)^n}}\mathrm{d}k}}+\int_{\lceil k_1\rceil}^{\infty}{k^{-\gamma}e^{\frac{(1-\theta)\lambda}{\theta\rho}\phi_{\infty}\cdot\sum_{n=0}^{\infty}{
\left(\frac{1}{\theta\rho k}\right)^n}}\mathrm{d}k}\right)\nonumber\\
&\approx&1-(\gamma-1)\left(\int_1^{\lfloor k_1\rfloor}{k^{-\gamma}e^{-(1-\theta)\lambda k\phi_{\infty}}\mathrm{d}k}+\int_{\lceil k_1\rceil}^{\infty}{k^{-\gamma}e^{\frac{(1-\theta)\lambda}{\theta\rho}\phi_{\infty}}\mathrm{d}k}\right)\nonumber\\
&=&1-(\gamma-1)\left(x_1^{\gamma-1}\left(\Gamma(-\gamma+1,x_1)-\Gamma(-\gamma+1,x_1 \lfloor k_1\rfloor)\right)+\frac{\lceil k_1\rceil^{-\gamma+1}}{\gamma-1}e^{x_2}\right)\nonumber\\
&\approx&A_1 x_1+A_2 x_2,\label{eq:a8}
\end{IEEEeqnarray}
where $x_1=(1-\theta)\lambda\phi_{\infty}$, $x_2=x_1/\theta\rho$, $A_1=\left((\theta\rho)^{\gamma-2}-1\right)\frac{\gamma-1}{\gamma-2}$, $A_2=-\left(\theta\rho\right)^{\gamma-1}$, and $\theta\rho k_1=1$. Note that if $k<\lfloor k_1\rfloor$, then $\theta\rho k<1$, and if $k>\lceil k_1\rceil$, then $\theta\rho k>1$.

Moreover, we write Eq.~\ref{eq:a8} as
\begin{equation}
\mathcal{P}=\left(A_1+A_2 x^{-1}\right)(1-\theta)\lambda\phi_{\infty},\label{eq:a9}
\end{equation}
where $x=\theta\rho$. Note that in the calculation of Eq.~\ref{eq:a9} and the following Eq.~\ref{eq:a91}, we use the Taylor expansion of $\Gamma(s,z)$, i.e., Eq.~\ref{eq:c3}. Thus Eqs.~\ref{eq:a9} and \ref{eq:a91} make sense only if $\gamma\notin \mathbb{N}$. The cases of $\gamma\in\mathbb{N}$ should be discussed case by case.

By multiplying Eq.~\ref{eq:15b} with $q(k)$, summing over $k$ and integrating over $t$, we get
\begin{equation*}
\frac{\mathrm{d}\phi(t)}{\mathrm{d}t}=\alpha-\alpha\langle\langle e^{-\lambda k\phi_k(t)}\rangle\rangle-\beta\phi(t).
\end{equation*}
As $t\rightarrow\infty$, we have $\mathrm{d}\phi/\mathrm{d}t\rightarrow0$. Then we get
\begin{IEEEeqnarray}{rCl}
\phi_{\infty}\approx\frac{\alpha}{\beta}\left(1-(\gamma-2)\int_1^{\infty}{k^{-\gamma+1}e^{-\lambda k\phi_k(\infty)}\mathrm{d}k}\right),\nonumber
\end{IEEEeqnarray}
which is similar to Eq.~\ref{eq:a7}. Hence we use the same technique and obtain
\begin{IEEEeqnarray}{rCl}
\phi_{\infty}&\approx&\frac{\alpha}{\beta}\left(B_1 x_1+B_2 x_1^2+B_3 x_2\right)\nonumber\\
&=&\left(B_1+B_2(1-\theta)\lambda\phi_{\infty}+B_3 x^{-1}\right)(1-\theta)\rho\phi_{\infty},\nonumber
\end{IEEEeqnarray}
where $x_1$, $x_2$ and $x$ have the same meaning as those in Eq.~\ref{eq:a8}, and $B_1=\left((\theta\rho)^{\gamma-3}-1\right)\frac{\gamma-2}{\gamma-3}$, $B_2=\left((\theta\rho)^{\gamma-4}-1\right)\frac{\gamma-2}{2(\gamma-4)}$, $B_3=-\left(\theta\rho\right)^{\gamma-2}$.
Then we have
\begin{equation}
(1-\theta)\lambda\phi_{\infty}=\frac{1-(1-\theta)\rho B_1-(1-\theta)\rho B_3 x^{-1}}{(1-\theta)\rho B_2}.\label{eq:a91}
\end{equation}
Given Eqs.~\ref{eq:a9} and \ref{eq:a91}, we get
\begin{IEEEeqnarray}{rCl}
\mathcal{P}=\frac{p_1(\theta)}{p_2(\theta)},\label{eq:a10}
\end{IEEEeqnarray}
where $\gamma$ is fixed, $p_1(\theta)$ and $p_2(\theta)$ are polynomials of $\theta$, and their coefficients are polynomials of $\rho$.

(ii) $\gamma=3$:

The approximation techniques used here are similar to those in case (i) except making use of Eq.~\ref{eq:c9} instead of Eq.~\ref{eq:c3}. The conclusion of this case is the same as that of case (i). For brevity, we omit the detailed proof here.
\section{Proof of Theorem 6}
We have proved the first part of the theorem. For the remaining part, by the deductions similar to those in Corollary~\ref{cor1} (see Appendix F), the maximum eigenvalue $\lambda_m$ of $\boldsymbol{L}$  equals approximately to $\alpha\lambda(1-\theta)\langle k^2\rangle/\langle k\rangle$. According to Definition~3, we get Eq.~\ref{eq:16}.
\section{Proof of Theorem 7}
Similarly to Eq.~\ref{eq:10}, we can write Eq.~\ref{eq:17b} as
\begin{equation}
\frac{\mathrm{d}\boldsymbol{a}(t)}{\mathrm{d}t}=\boldsymbol{L}(t)\boldsymbol{a}(t),\label{eq:19}
\end{equation}
where the Jacobian matrix $\boldsymbol{L}(t)=\{L_{kk'}(t)\}$ is defined by
\begin{equation*}
L_{kk'}(t)=\alpha(t)\lambda(t) kP(k'\mid k).
\end{equation*}

Note that $\boldsymbol{L}(t)=\alpha(t)\lambda(t)\boldsymbol{C}$, where $\boldsymbol{C}$ is the connectivity matrix defined in Theorem~\ref{thm4}. Thus, $\boldsymbol{L}(t)$ is commutative, {i.e.}, $\boldsymbol{L}(t_1)\boldsymbol{L}(t_2)=\boldsymbol{L}(t_2)\boldsymbol{L}(t_1)$ for all $t_1,t_2\in(0,\infty)$. By defining
\begin{equation*}
\hat{\boldsymbol{L}}(t)\triangleq\int_0^t \boldsymbol{L}(x)\mathrm{d}x,
\end{equation*}
we get $\hat{\boldsymbol{L}}(t_1)\boldsymbol{L}(t_2)=\boldsymbol{L}(t_2)\hat{\boldsymbol{L}}(t_1)$. Based on these facts, we can easily get
\begin{equation*}
\frac{\mathrm{d}\hat{\boldsymbol{L}}^n(t)}{\mathrm{d}t}=n\hat{\boldsymbol{L}}'(t)\hat{\boldsymbol{L}}^{n-1}(t).
\end{equation*}

We claim that $\boldsymbol{a}(t)=\exp{\hat{\boldsymbol{L}}(t)}\cdot\boldsymbol{a}(0)$ is the solution of Eq.~\ref{eq:19}. To prove this, by taking the derivative of $\boldsymbol{a}(t)$, we have
\begin{IEEEeqnarray}{rCl}
\frac{\mathrm{d} \boldsymbol{a}(t)}{\mathrm{d}t}&=&\frac{\mathrm{d}}{\mathrm{d}t}\left(\sum_{n=0}^{\infty}{\frac{\hat{\boldsymbol{L}}^n(t)}{n!}}\cdot\boldsymbol{a}(0)\right)=
\sum_{n=0}^{\infty}{\hat{\boldsymbol{L}}'(t)\frac{\hat{\boldsymbol{L}}^n(t)}{n!}}\cdot\boldsymbol{a}(0)\nonumber\\
&=&\hat{\boldsymbol{L}}'(t)\exp{\hat{\boldsymbol{L}}(t)}\cdot\boldsymbol{a}(0)=\boldsymbol{L}(t)\boldsymbol{a}(t),\nonumber
\end{IEEEeqnarray}
which is exactly Eq.~\ref{eq:19}.

Since $\boldsymbol{C}$ is a real symmetric matrix, it can be diagonalized by an orthogonal matrix. Furthermore, $\boldsymbol{C}$ is nonsingular according to the proof of Corollary~\ref{cor1} (see Appendix F). Thus it is similar to the identity matrix, {i.e.}, there exists an orthogonal matrix $\boldsymbol{P}$ such that $\boldsymbol{P}^{-1}\boldsymbol{C}\boldsymbol{P}=\boldsymbol{I}$. Therefore, we can rewrite the solution of Eq.~\ref{eq:19} as
\begin{IEEEeqnarray}{rCl}
\boldsymbol{a}(t)&=&\exp{\hat{\boldsymbol{L}}(t)}\cdot\boldsymbol{a}(0)=\exp\left(\int_0^t{\boldsymbol{L}(x)\mathrm{d}x}\right)\boldsymbol{a}(0)\nonumber\\
&=&\exp\left(\int_0^t{\alpha(x)\lambda(x)\mathrm{d}x}\cdot\boldsymbol{C}\right)\boldsymbol{a}(0)\nonumber\\
&=&\exp\left(\int_0^t{\alpha(x)\lambda(x)\mathrm{d}x\cdot\boldsymbol{P}\boldsymbol{I}\boldsymbol{P}^{-1}}\right)\boldsymbol{a}(0)\nonumber\\
&=&\sum_{n=0}^{\infty}\frac{1}{n!}\left(\boldsymbol{P}\left(\int_0^t\alpha(x)\lambda(x)\mathrm{d}x\cdot\boldsymbol{I}\right)\boldsymbol{P}^{-1}\right)^n\boldsymbol{a}(0)\nonumber\\
&=&\boldsymbol{P}\exp\left(\int_0^t{\alpha(x)\lambda(x)\mathrm{d}x}\cdot\boldsymbol{I}\right)\boldsymbol{P}^{-1}\boldsymbol{a}(0)\nonumber\\
&=&\exp\left(\int_0^t{\alpha(x)\lambda(x)\mathrm{d}x}\right)\boldsymbol{P}\boldsymbol{I}\boldsymbol{P}^{-1}\boldsymbol{a}(0)\nonumber\\
&=&\exp\left(\int_0^t{\alpha(x)\lambda(x)\mathrm{d}x}\right)\boldsymbol{a}(0).\label{eq:20}
\end{IEEEeqnarray}
Together with Eq.~\ref{eq:19}, we finally obtain Eq.~\ref{eq:18}.

\section{Proof of Corollary 2}
According to Theorem~\ref{thm7}, we have
\begin{IEEEeqnarray}{rCl}
\frac{\mathrm{d}a(t)}{\mathrm{d}t}&=&\frac{\mathrm{d}}{\mathrm{d}t}\sum_k{P(k)a_k(t)}=\sum_k{P(k)\frac{\mathrm{d}a_k(t)}{\mathrm{d}t}}\nonumber\\
&=&\sum_k\left(P(k)\alpha(t)\lambda(t)\exp\left(\int_0^t{\alpha(x)\lambda(x)\mathrm{d}x}\right)\sum_{k'}{k P(k'\mid k)a_{k'}(0)}\right)\nonumber\\
&=&\alpha(t)\lambda(t)\exp\left(\int_0^t{\alpha(x)\lambda(x)\mathrm{d}x}\right)C.\nonumber
\end{IEEEeqnarray}

\end{appendices}

\small
\bibliographystyle{ieeetr}
\bibliography{paper}
\end{document}